\pdfoutput=1
\documentclass[conference]{IEEEtran}
\IEEEoverridecommandlockouts

\usepackage{marvosym}
\usepackage{url}
\usepackage{pifont}
\usepackage{xspace}
 \usepackage{amssymb}
\usepackage{amsmath,amsfonts} 
\usepackage{stfloats}
\usepackage{algorithmic}
\usepackage{graphicx}
\usepackage{textcomp}
\usepackage{xcolor}
\usepackage{multirow}
\usepackage{makecell}
\usepackage[linesnumbered,ruled,lined]{algorithm2e}
\def\BibTeX{{\rm B\kern-.05em{\sc i\kern-.025em b}\kern-.08em
    T\kern-.1667em\lower.7ex\hbox{E}\kern-.125emX}}

\usepackage[font=small, labelfont=bf]{caption}

\newcommand{\sys}{AVMiner\xspace}
\newcommand{\AVCLASS}{AV{\scriptsize CLASS}\xspace}
\newcommand{\AVCLASSTWO}{AV{\scriptsize CLASS}2\xspace}

\newcommand{\cmark}{\ding{51}}

\begin{document}

\title{\sys: Expansible and Semantic-Preserving Anti-Virus Labels Mining Method}


\author{Ligeng Chen, Zhongling He, Hao Wu, Yuhang Gong, Bing Mao, Nanjing University}

\author{
\IEEEauthorblockN{Ligeng Chen\IEEEauthorrefmark{2}\IEEEauthorrefmark{4}\Letter, Zhongling He\IEEEauthorrefmark{2}, Hao Wu\IEEEauthorrefmark{2}, Yuhang Gong\IEEEauthorrefmark{2} and Bing Mao\IEEEauthorrefmark{2}} 
  
\IEEEauthorblockA{\IEEEauthorrefmark{2}National Key Laboratory for Novel Software Technology, Nanjing University, Nanjing, China \\ 
\{chenlg, zhe, gyh\}@smail.nju.edu.cn, \{hao.wu, maobing\}@nju.edu.cn}

\IEEEauthorblockA{\IEEEauthorrefmark{4}Asiainfo Security Technologies Co.,ltd}
}



\maketitle

\begin{abstract}

With the increase in the variety and quantity of malware, there is an urgent need to speed up the diagnosis and the analysis of malware. Extracting the malware family-related tokens from AV (Anti-Virus) labels, provided by online anti-virus engines, paves the way for pre-diagnosing the malware.
Automatically extract the vital information from AV labels will greatly enhance the detection ability of security enterprises and equip the research ability of security analysts. Recent works like \AVCLASS and \AVCLASSTWO try to extract the attributes of malware from AV labels and establish the taxonomy based on expert knowledge.
However, due to the uncertain trend of complicated malicious behaviors, the system needs the following abilities to face the challenge: preserving vital semantics, being expansible, and free from expert knowledge.
In this work, we present \sys, an expansible malware tagging system that can mine the most vital tokens from AV labels. \sys adopts natural language processing techniques and clustering methods to generate a sequence of tokens without expert knowledge ranked by importance. \sys can self-update when new samples come. Finally, we evaluate \sys on over 8,000 samples from well-known datasets with manually labeled ground truth, which outperforms previous works.

\end{abstract}




\begin{IEEEkeywords}
AV Labels, Expansible, Natural Language Processing
\end{IEEEkeywords}


\section{Introduction}

According to the report of AV-TEST\cite{avtest}, 130 million new viruses were detected in the past year (4.1 viruses per second on average). Rapid mutating malware have a great impact on daily lives, causing huge losses to people's properties. To enhance the detection ability of vendors for the mutants of malware, it must have a sufficient understanding of these malicious samples. So it is the best way to accumulate the complete information of these malicious samples and the detailed analysis results. But the analysis needs to go through many procedures to verify the platform, programming languages, malicious behaviors, and the malware families at the end. Each procedure of the verification requires much expert knowledge to extract features equipped with rules. So it is urgent to come up with an efficient way to accelerate the procedure for the newly coming malwares.



Hence, precisely tagging a suspicious malware in advance will significantly help security enterprises to archive the malware by different families and assist security analysts in further analyze the malicious behaviors and patch the vulnerabilities. Some online services like VirusTotal\cite{virustotal} provide malware labels detected by various anti-virus engines, which are called AV labels. The naming way of AV labels follows the original intention of MAEC (Malware Attribute Enumeration and Characterization)\cite{MAEC}, aiming to propose a structured representation with high-fidelity information about the attributes of malware. 
However, since the labels are produced by different vendors independently, they are always inconsistent with each other on the judgment of characteristics (i.e., malicious or benign), descriptions (e.g., malware class, property, and behavior), and so on. Previous works have located the label inconsistency problem. What's more, AV-Meter\cite{mohaisen2014avmeter} empirically studies the correctness and inconsistency of malware tagging on manually labeled datasets. Even well-known vendors can not always perform well. Based on the observation that tags change over time, Zhu et al.\cite{zhu2020measuring} thoroughly research and show that AV labels would flip over time, speculating AV vendors produce strongly correlated labels.

Even though the massive AV labels are noisy, some works\cite{sebastian2016avclass,avclass++,sebastian2020avclass2} demonstrate that the ground-truth of tags is hidden in the given labels, which can be automatically extracted with rules built by expert knowledge.

Unfortunately, due to the unpredictable trend of malware, the complicated relationships between malicious behaviors, and the inconsistent description of AV labels, we are still facing the following challenges of automatically extracting the correct labels for the malware samples. 
\ding{182} Complex relationships of malicious behaviors and vendor naming rules are challenging to be exhausted by the limited rules. Is it possible to extract the labels and reunite the relationships between them \textit{\textbf{Without Expert Knowledge}}?
\ding{183} Although different AV vendors produce inconsistent labels, but the labels shows an inherent relation. According to the different detection mechanisms (e.g., static analysis, sandbox execution), they focus on different aspects of the malware. A single label may only describe a malicious behavior profile, but breaking up and regrouping the massive AV labels may give a full picture of the sample. So how to \textit{\textbf{Preserve Vital Semantics}} rather than directly follow the principle of the minority obeying the majority?
\ding{184} The arms race between hackers and security researchers is increasingly fierce. Researchers are unable to predict the characteristics of the new kind of viruses before their appearance, let alone categorize them. Some new malware, e.g., HeartBleed\cite{heartbleed}, EternalBlue\cite{eternalblue}, and Vjworm\cite{vjworm}, cannot be classified into any malware type. So how to make the system \textit{\textbf{Expansible}} to adapt to the development of malware? 

To deal with the problems mentioned above, we present a novel malware tagging method called \sys (\textbf{A}nti-\textbf{V}irus \textbf{Miner}), which can adapt to new data with no human effort. 
\ding{172} \sys adopts state-of-the-art NLP (Natural Language Processing) techniques and adaptive clustering techniques to extract the labels for the malware, rather than establish excessive rules with massive expert knowledge. The system will tokenize, generalize and extract the key label(s) and orchestrate them according to their relations \textit{\textbf{Without Expert Knowledge}}.
\ding{173} By utilizing machine learning techniques, we can be freed from extracting endless rules. We try to \textit{\textbf{Preserving the Vital Semantics}} as much as possible, based on the principle of co-occurrence frequency. The extracted labels will be reunited to a relation graph, which will give a full picture of the malware to the users.
\ding{174} Due to the system design, \sys requires almost no expert knowledge. With the mutation of malware, it is able to update self-adaptively to deal with coming malicious samples and its AV labels. So the \textit{\textbf{Expansibility}} of the system can meet the community's need.

We evaluate \sys on over 8,000 samples with manually labeled ground truth and compare it with state-of-the-art malware label extraction method. \sys achieves an accuracy of 93.5\% when outputs 3 labels, and 97.9\% when outputs 5 labels, separately higher than \AVCLASS and \AVCLASSTWO by about 7\% and 9\%. Also, we evaluate \sys on about 2 million samples without ground truth, which can give the output as long as the AV labels are given. But \AVCLASS and \AVCLASSTWO fail to output the result for 21.8\% and 10.6\% of samples.
\sys shows great robustness and expansibility on the diverse distribution of malicious samples.



The main contributions of our work is as follows:

\begin{enumerate}
	\item We propose a novel malware label generation method by mining the malware label from the massive AV labels collected from security communities to accelerate the malware diagnosing procedure. 
	\item We design a new tool \sys to effectively generate the tokens that express malware's security semantics from the multi-source and noisy AV labels. \sys works without expert knowledge and can self-adaptively update with the newly come samples.
	\item We evaluate the \sys on 2 manually labeled datasets and measure it on about 2 million malicious samples without ground truth. Experimental results show that \sys behaves better than the existing works.
\end{enumerate}


\section{Motivation and Definition}

\subsection{Motivation}

The accumulation of malicious samples is significant for security enterprises and laboratories. These samples record the malicious behaviors of hackers, of which the trend even can be indicated. However, carefully analyzing and archiving the malware set up a barrier due to the limited information provided, consuming a great amount of time and expert knowledge. The ordinary procedure of analyzing malware and confirming the family of malware is presented as follows.

\begin{figure}[!h]
    \centering
    \includegraphics[width=8.5cm]{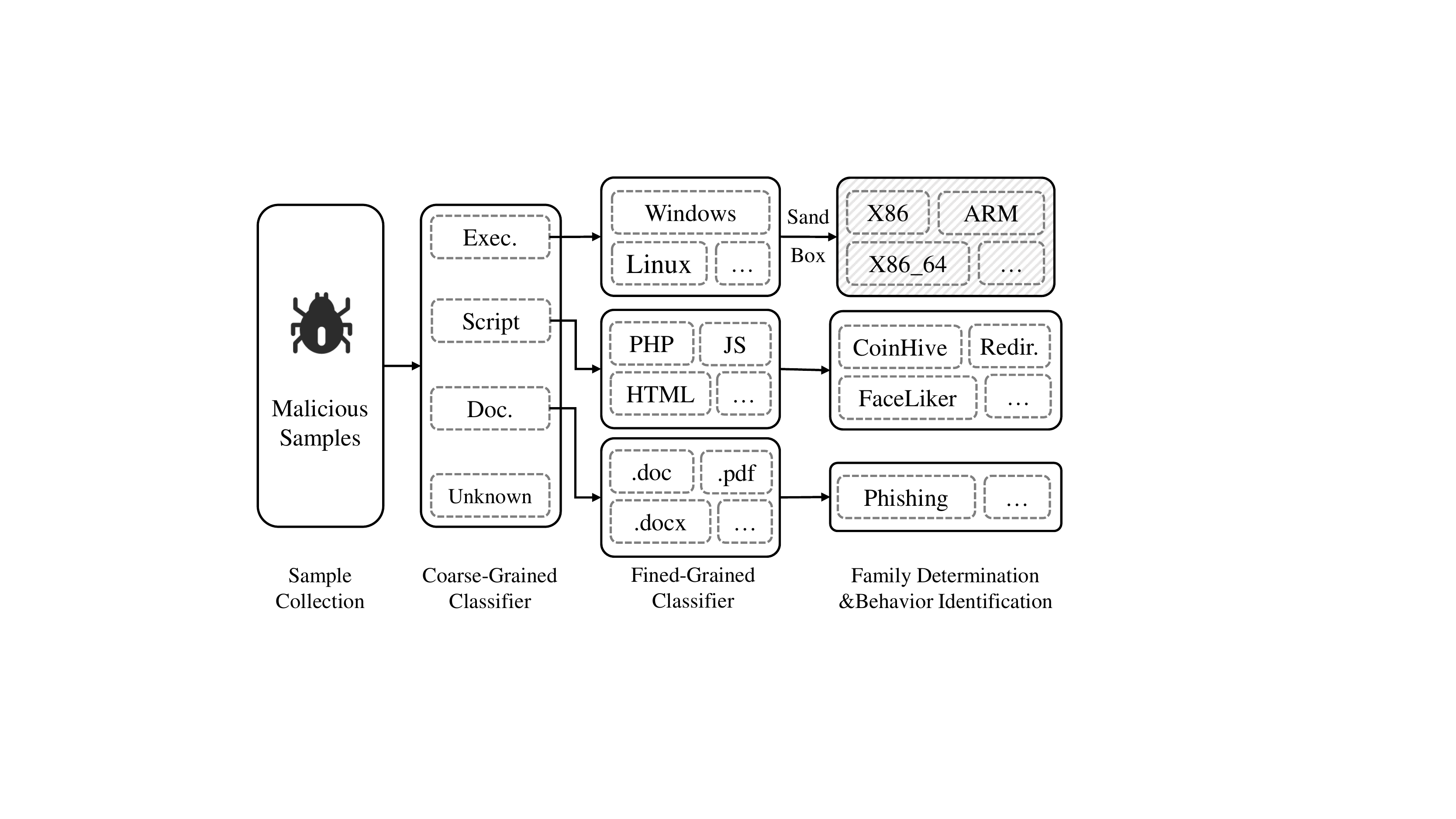}
    \caption{Industrial pipeline for security enterprises to archive malwares.}
    \label{fig:malwarearchive}
\end{figure}

As shown in Figure~\ref{fig:malwarearchive}, the pipeline mainly contains 4 stages. Firstly, the wide range of malware collection (e.g., publicly accessible website or online security community) sets up the repository. According to the execution type of malicious samples, they will be classified into 4 kinds (i.e., executable, script, document, and unknown type) under a coarse-grained way. Then, the samples will be classified in a fine-grained way according to the programming languages and platforms. At last, they will be separately put into the procedures of static analysis or dynamic analysis to identify the malicious behavior and determine the malware family.

\vspace{-8pt}
\vspace{-8pt}

However, even though we can successfully distinguish the execution type of malware, it still lacks automatic methods for the rest of the stages. Due to the obfuscation techniques for scripts and the insidious behaviors of executables, the malware that cannot be accurately identified the family requires a lot of trials, which is time-consuming and labor-intensive (each executable file requires about 180s to execute in the sandbox at one time, and each script requires about 60s to analysis at one time). If we can get the precise family of the malware with extra information in advance, it can extremely accelerate the archive procedure.



To validate our thoughts, we evaluate 100,000 malware sampled from real-world ranging from the year 2006 to 2020. By processing the results from VirusTotal (i.e., AV labels) of each malware, we rank the extracted tokens by customized weight, aiming to make the token representing the family of malware as front as possible. The result is shown in Table~\ref{tab:motivationresult}, compared with previous labeling methods.

\begin{table}[!h]
\centering
\small
\caption{Result of capturing the malware family via AV labels.}
\begin{tabular}{c|ccc}
\Xhline{1.2pt}
 & Recall@1 & Recall@3 & Recall@5 \\\Xhline{1.2pt}
\AVCLASS\cite{sebastian2016avclass}   & 54.0\%   & 54.0\%   & 54.0\%   \\
\AVCLASSTWO\cite{sebastian2020avclass2} & 45.5\%   & 71.7\%   & 75.0\%  \\
\sys  & 74.5\%   & 88.5\%   & 94.9\%   \\\Xhline{1.2pt}
\end{tabular}

\label{tab:motivationresult}
\end{table}

According to the table, column 2 to 4 denote the percentage of malware can be extracted the corresponding malware family from AV labels. That is, 74.5\% of malwares' family matches the first token produced by our method \sys, and it raises to 88.5\% and 94.9\% when the output range is expanded to top 3 and top 5. In other words, 74.5\% of malicious samples can be determined by the malware family by going through the industrial pipeline only once assisted by the first token produced by \sys, while the hit rate drops to 54.0\% when the labeling system is \AVCLASS. 

To conclude, equipped with a labeling system, the analysis component in archiving the malware even can be skipped. We just need to verify the malicious behavior of the malware with the extracted information about the malware family. Unfortunately, even the previous works give a rule-based solution by extracting the tokens from AV labels, but with the quick development of hack attacks and wide range of virus variance, rules will out of date one day. So we try to improve the labeling mechanism with the following needs.



\paragraph{\textbf{Free from Expert Knowledge}}
With the rapid development of information technology, there are growing security issues of information and software. To capture the characteristics of all kinds of malware as much as possible, vendors actively update their scanning rules. 

By extracting labels from AV labels, previous works are fully based on rules generated by expert knowledge to deal with the problem of miscellaneous labeling rules of different vendors and unpredictable behavior of malware.

\begin{table}[!h]
    \centering
    \small
    \caption{Amount of required rules extracted manually from previous works (\AVCLASS and \AVCLASSTWO).}
    \begin{tabular}{c|c|c}
    \Xhline{1.2pt}
         & \AVCLASS\cite{sebastian2016avclass} & \AVCLASSTWO \cite{sebastian2020avclass2} \\\Xhline{1.2pt}
    Tokenization & \# of Vendors &  \# of Vendors\\
    Token Filter & 417 &1,300\\
    Token Alias Merge& 559  &  1,138\\ \Xhline{1.2pt}
    \end{tabular}
    
    \label{tab:rulesofpreviousworks}
\end{table}

According to the Table~\ref{tab:rulesofpreviousworks} shown above, column 2 and 3 denotes the rule amount of \AVCLASS\cite{sebastian2016avclass} and \AVCLASSTWO\cite{sebastian2020avclass2}. Row 2 indicates the amount of vendor-customized rules to tokenize for each vendor. Row 3 and 4 denote the required rules to firstly filter generic or irrelevant tokens and then merge the alias tokens. It is obvious that much human effort and expert knowledge are needed to generate adequate rules and build such a system. 
Nevertheless, malware is evolving unpredictably, and the format of AV labels produced by the vendors may also change over time. Therefore, after a period of time, these systems with fixed rules may partially fail due to the change of inputs. To solve this problem, we need to come up with an intelligent method free from expert knowledge.

\begin{figure}[!h]
    \centering
    \includegraphics[width=8.5cm]{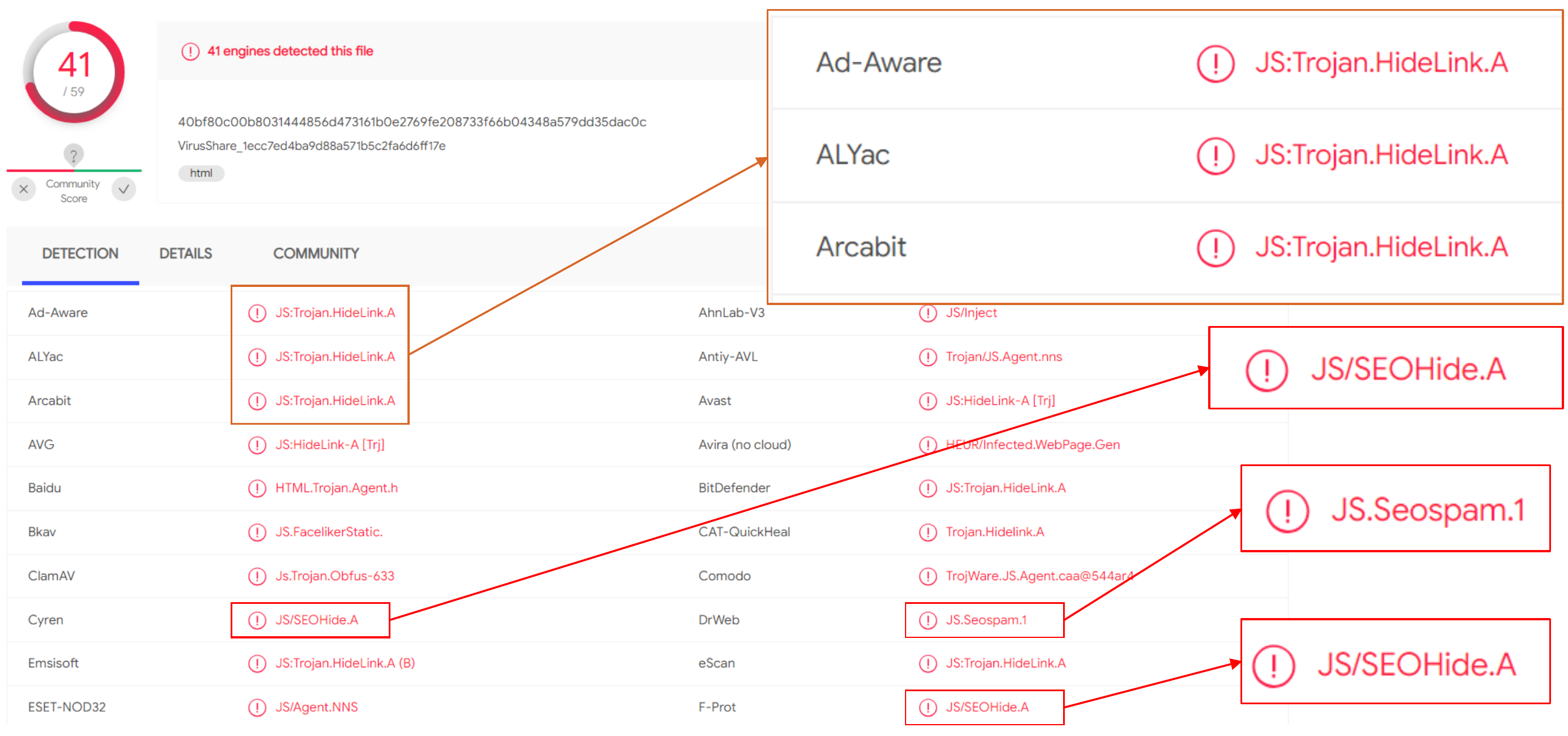}
    \caption{The web view of VirusTotal presenting the detection result of motivation example.}
    \label{fig:motivationexample}
\end{figure}

\paragraph{\textbf{Semantic Preserving}}
Once we get a malicious sample, we can put it to the online detection platform, such as VirusTotal\cite{virustotal}, to get preliminary feedback from its API, which includes results from some vendors. As shown in Figure~\ref{fig:motivationexample}, it is a report that consists of AV vendors' individual detection results for malware.

According to the results shown in the figure, we find that the first three vendors (i.e., Ad-Aware, ALYac, Arcabit) give exactly the consistent judgments \textit{JS:Trojan.HideLink.A}. 

It is easy to understand that they all believe that this malware is a \textit{Trojan} program in JS format, which has malicious behavior of \textit{hidding links} in the program. However, the three vendors in the three individual frames do not give the same detection results. They think it has a hidden link that is maliciously implanted in \textit{search engine optimization} (\textit{SEO} for abbreviation). Although previous works have studied the problem of label inconsistency, in fact, in most cases, the labels given by the vendors are not completely orthogonal. Instead, they describe a sample from a different granularity or describe different profiles, (e.g., infection targets, replication techniques, and stealth techniques). According to this example, \textit{SEO} is a more fine-grained description of \textit{hidelink}.

\begin{table}[!h]
\scriptsize
\centering
\caption{Comparison result of motivation example.}
\begin{tabular}{c|l}
\Xhline{1.2pt}
Tool & \multicolumn{1}{c}{Result}                        \\\Xhline{1.2pt}   
\AVCLASS\cite{sebastian2016avclass}                   & hidelink                                             \\\hline


\multirow{2}{*}{\AVCLASSTWO\cite{sebastian2020avclass2}} & hidelink,7$\|$html,5$\|$webpage,2$\|$network,2$\|$jswebinject,2 \\
                          & infected,2$\|$execdownload,2$\|$downloader,2               \\\hline

\multirow{2}{*}{\sys}  & hidelink,12$\|$seohide,3$\|$trojan,11$\|$redirector,3    \\
                          & infected,2$\|$webpage,2
                          \\\Xhline{1.2pt}               
\end{tabular}

\label{tab:rescomparisonofMotivation}
\end{table}

According to our observation, we find that previous tools seem too rough to tag the malware from AV labels. \AVCLASS aims to merge the massive labels into a single label to represent the class of the malware, while \AVCLASSTWO aggregates the alias labels and outputs them with a pre-set taxonomy. This kind of operation may abandon some meaningful tags with low appearance frequency but also make the labeling system not adaptable enough.

To verify our thought, we try to process the sample with different methods (i.e., \AVCLASS, \AVCLASSTWO, and our system \sys) and get the results shown in Table~\ref{tab:rescomparisonofMotivation}. All these tools process the AV labels produced by VirusTotal related to the example malware. The symbol $\|$ is used to separate the labels. And each tuple represents the number of the current label calculated by the tool. For instance, the first tuple 'hidelink, 7' produced by \AVCLASSTWO stands for 7 vendors labeling the malware as hidelink.

The reason for the inconsistency between our system \sys and \AVCLASSTWO on the amount of \textit{hidelink} is that \AVCLASSTWO deletes the labels suspected of using the same detection mechanism or suspected to belong to the same manufacturer. So the number of \textit{hidelink} is reduced sharply, but we do not deal with this situation. Therefore, we need to design the system with the consideration of preserving the words that contain the secondary semantics which describe the profile of malware.

\paragraph{\textbf{Expansibility}}
Due to the system design, \AVCLASS does not consider the expansibility of the tool. Even \AVCLASSTWO considers the issue and is equipped with an open taxonomy, which still relies on thousands of rules. Expected, even \AVCLASSTWO fail to label about 20\% of the samples in 2020, do not speak of \AVCLASS fails 40\%.

According to the 3 aspects mentioned above, we try to propose a malware tagging tool that is easy-to-use for both users (with less expert knowledge) and security analysts (with much expert knowledge), even security enterprises. 
Due to the limited introduced prior knowledge, we can not precisely narrow down the output into one or two words but produce a set of candidate keywords to describe the malicious sample.


\subsection{Problem Definition}

For clarity, we illustrate and give a clear definition of several commonly used words that are highly related to our problems and system design. 

\begin{figure}[!h]
    \centering
    \includegraphics[width=8.5cm]{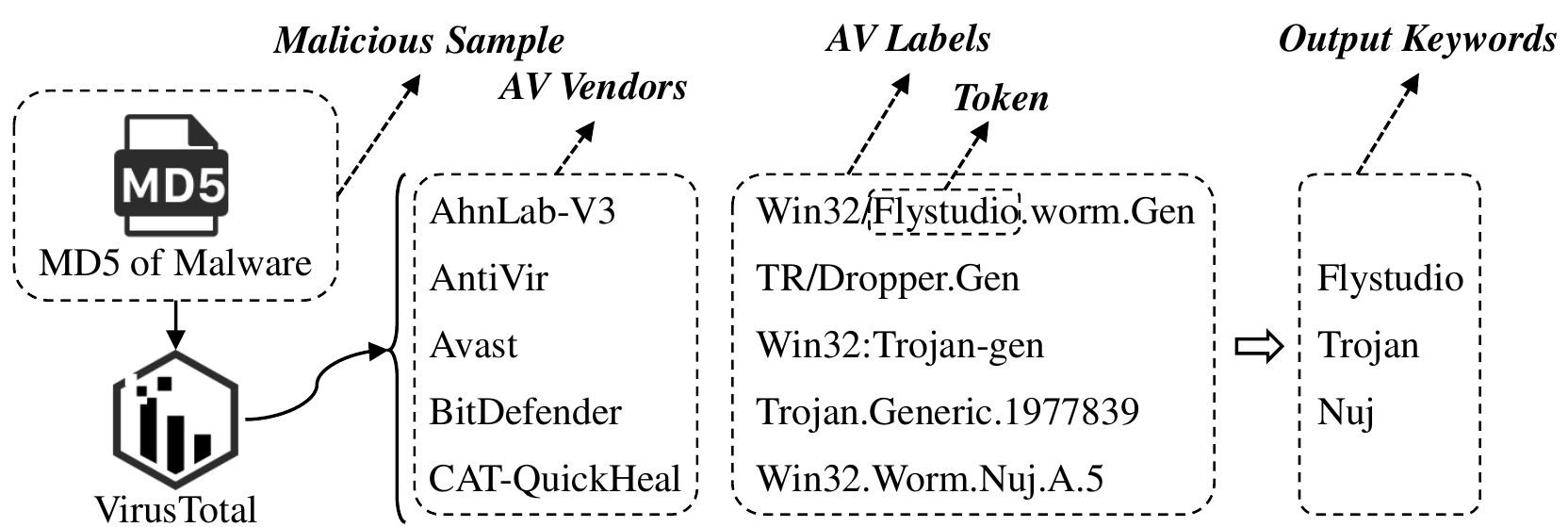}
    \caption{The descriptions of commonly used words in the article.}
    \label{fig:clarity}
\end{figure}

As shown in Figure~\ref{fig:clarity}, we firstly input the MD5 code or the executables of the \textit{Malicious Samples} to VirusTotal. And various \textit{AV Vendors} output the \textit{AV Labels} according to their detecting methods. Each element separated by different punctuations in the \textit{AV Lables} is referred as a \textit{Token}. Our system tries to process the \textit{AV Labels} to a single or a set of consistent \textit{Keywords}, aiming to mine the most related and precise description of the malicious samples for the users.

Some formal descriptions are given as follows,

\textbf{Input:} A set of malicous samples is defined as,

\begin{align}
    Mal= \{mal_1, mal_2, ..., mal_i\}
\end{align}

\textbf{Intermedieate output:} A set of corresponding AV labels $Label_{V}$ and a set of tokens extracted from AV labels are defined as,

\begin{align}
    Label_{V}= \{label_{v1-1}, label_{v_2-1}, ..., label_{v_j-i}\}, \\Token = \{T_1, T_2, ..., T_n\} 
\end{align}


where $label_{v_j-i}$ denotes the label produced by $j^{th}$ vendor to describe the $i^{th}$ sample. $T_n$ denotes the $n^{th}$ token in the token corpus. After being scanned by VirusTotal, the malicious samples are tagged with AV labels, which are denoted as,

\begin{align}
    L(mal_i)= \{label_{v1-i}, label_{v2-i},... ,label_{v_j-i}\}
\end{align}

\textbf{Output:} A set of output keywords is defined as,

\begin{align}
    Keyword=\{kwd_{top_1-i}, kwd_{top_2-i}, ..., kwd_{top_k-i}\}
\end{align}

where $kwd_{top_k-i}$ denotes the extracted keyword ranked $k^{th}$ to describe the $i^{th}$ sample.


\section{System Design}

\subsection{Overview}

\begin{figure}[!h]
    \centering
    \includegraphics[width=8.5cm]{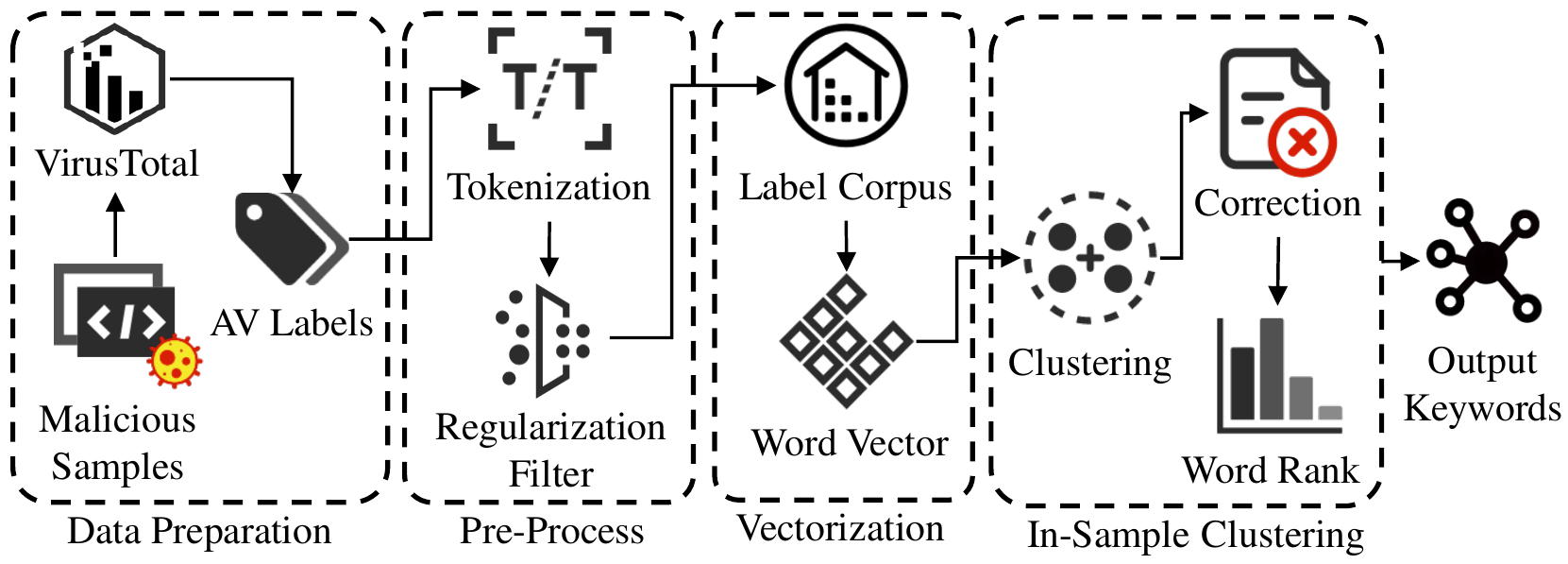}
    \caption{Workflow of \sys, containing 5 Parts: Data Preparation, Pre-Process, Vectorization, In-Sample Clustering, and Output.}
    \label{fig:workflow}
\end{figure}

\begin{figure}[!h]
    \centering
    \includegraphics[width=8.5cm]{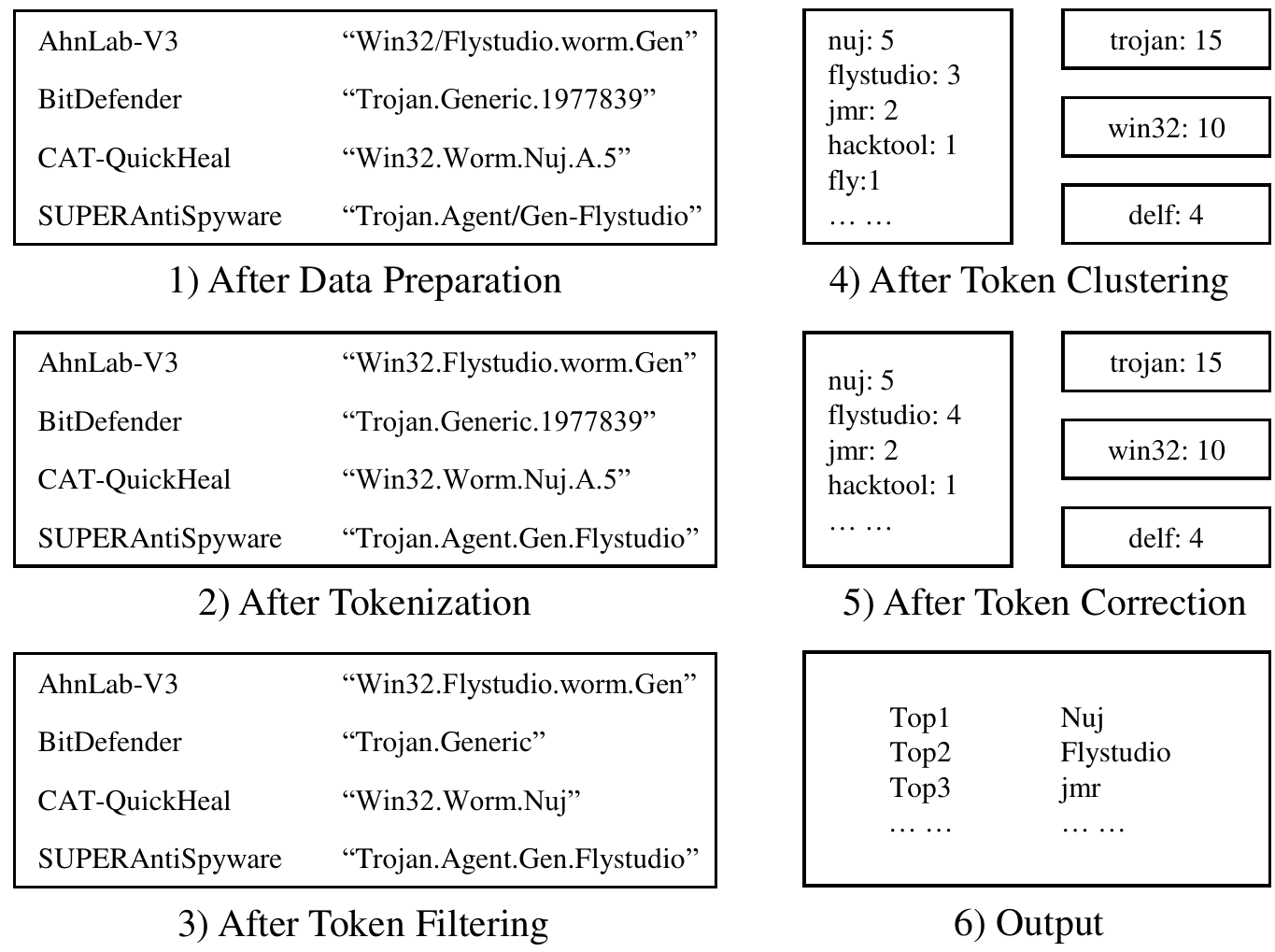}
    \caption{Running example. Left side of step 1) to 3) denotes the name of AV vendors, right side denotes the AV labels and processed tokens. Step 4) and 5) contains different clustered tokens and their amount. Step 6) outputs the ranked tokens.}
    \label{fig:runningexample}
\end{figure}

In this section, we provide an informal overview of our method \sys on an illustrative example as shown in Figure~\ref{fig:runningexample}. According to Figure~\ref{fig:workflow}, \sys mainly contains five parts: \textit{data preparation}, \textit{pre-processing}, \textit{vectorization},  \textit{in-sample clustering} and \textit{keywords output}, each of which will be introduced in detail in the following subsections.

\subsection{Data Preparation}

The Data Preparation phase is shown on the left side of Figure~\ref{fig:workflow}. In this phase, we first collect the malware' detected results from the VirusTotal (denoted as AV label). Each AV label contains a sequence of keywords indicating the detected result from each vendor assembled on VirusTotal, as shown in Figure~\ref{fig:runningexample}-1. These AV labels describe the possible behavior, attributes, and categories of a malicious sample (e.g., "Win32\/Flystudio.worm.Gen" produced by \textit{AhnLab-V3}, "Win32" denotes the platform, "Flystudio" denotes the category, "worm" denotes the behavior). In the following phases, \sys will process the AV labels into an intermediate representation and extract the keywords best describing the samples (i.e., "Flystudio" in the aforementioned example) for the users.

To free from summarizing excessive rules and make the best use of advanced techniques, our \sys is based on the statistical results and equipped with machine learning methods. To deal with the problem of cold start with zero knowledge, we gather a considerable amount (more than 8,000 in the evaluation) of samples to set up the corpus in the following phase. When new samples come, they can be directly added to the system and make the tool adaptive to the unknown kinds of malware. Abundant and diverse datasets of malicious samples can not only make our systems more inclusive for a different kind of virus but also enable us to build a virus-related knowledge graph in the future.




\subsection{Pre-Processing}

The pre-processing phase consists of two parts: tokenization and token filter. 
The tokenization aims to split the AV labels into tokens for the same token merging and related tokens clustering. The token filtering aims to filter out the tokens with a few semantics and prevent them from blocking the subsequent stages.  

\paragraph{Tokenization}
The AV labels are sequences of tokens and highly vendor-dependent. For example, in the face of a Trojan sample which can be classified as \textit{delf}, \textit{Emsisoft} describes it as \textit{Trojan-Dropper.Delf!IK}, but it is described as \textit{Trojan.DL.Delf!qqcViDnxCRM} by \textit{VirusBuster}.
The various strategies of AV vendors to tag the malware leads to the AV labels appearing with quite different formats but with even the same semantics.
So it is vital to split the AV labels to find the associated parts between the inconsistent labels. What's more, the trend of malware development is unpredictable, and AV vendors may update tagging strategies over time. But it takes time to customize rules that cause the rules to always out-of-date.
So customizing a set of constant rules is hard. To address the challenges mentioned above, we firstly tokenize the AV labels into the smallest units, called tokens, which are separated by some specific punctuations in AV labels. 
Specifically, the compound words in our scenario should be preserved, because these words usually have specific semantics in a particular context.
To clearly present the result, we replace all the punctuations with a unified separator, denoting the tokenization operation. 
According to the example in Figure~\ref{fig:runningexample}-2, the AV label \textit{Win32/Flystudio.worm.Gen} produced by the vendor \textit{AhnLab-V3} is related to a malware sample. We process the AV label into \textit{Win32.Flystudio.worm.Gen}, containing 4 tokens, separated by a dot, leaving alone the compound word \textit{Flystudio} and the abbreviation \textit{Gen} (representing for word \textit{generic}).

\paragraph{Token filter}
Performing tokenization may produce a large number of redundant and meaningless tokens, so we need to remove them, aiming to minimize the influence of noise. For example, according to Figure~\ref{fig:runningexample}-2), AV label "Win32.Worm.Nuj.A.5" produced by \textit{CAT-QuickHeal} comprises two tokens \textit{A} and \textit{5}, which seem with low-security semantics for the malware and may only be used as a malware identifier or only as a serial number.
Although we find this kind of tokens is often appear in some fixed location, it cannot be removed directly due to the following reason. The token number of each AV label is quite different, ranging from 1 to 6. Even the AV labels produced from the same vendor may not be aligned. For example, the token number of AV labels produced by \textit{BitDefender} is ranging from 3 to 5. 



Through a detailed investigation of the collected AV labels, we find the there are some key differences between the meaningful tokens and the meaningless ones. \textit{First}, the AV labels from the same vendor have almost the same formal structures. For example, the first token of \textit{BitDefender} is usually a feature description (e.g., generic, trojan), the second is usually a file type description (e.g., JS), and the last token is usually a number or identifier. Although the length may vary depending on the description, the same type of tokens is generally in the same relative position. \textit{Second}, the meaningful tokens usually have a high repetition frequency, and they are often used to describe an attribute, family, class, etc, while the meaningless tokens are only used as an identifier with low repetition frequency. So the token position and the token distribution light the way for us to effectively filter the meaningless tokens. We come up with a strategy called \textbf{\textit{Unique Tokens Fade Away}}. According to the observation, the same relative token position (in order or reverse) of AV labels produced by the vendor always containing the same type of tokens as mentioned. And the meaningless tokens are unique in the same relative position. When we count the proportion of unique tokens in all the tokens, this may lead to a clear distinction. Here we define the unique index $\sigma$ as,

\begin{align}
    \sigma(i,i_{max},v_t,f) =
    \begin{cases}
    \frac{T_{u}(i,v_t)}{T_{a}(i,v_t)},f=1\\
    \frac{T_{u}(i_{max}-i,v_t)}{T_{a}(i_{max}-i,v_t)},f=0
    \end{cases}
\end{align}

where $\sigma$ indicates that the proportion of unique tokens (i.e., tokens appear only once) in all the tokens. The calculation of $\sigma$ is restricted to the same relative position of the same vendor. The parameters of $\sigma$ are defined as follows: $i$ denotes the $i^{th}$ position of the AV label, $i_{max}$ denotes the longest token sequence in the same vendor, $v_t$ denotes the $t^{th}$ AV vendor, $f$ denotes a flag whether we count the token orderly (f=1) or reversely (f=0). $T_{u}(i,v_t)$ denotes the amount of the unique tokens in the $i^{th}$ position from the $t^{th}$ AV vendor, $T_{a}(i,v_t)$ denotes the amount of all the tokens in the $i^{th}$ position from the $t^{th}$ AV vendor.


\begin{figure}[!h]
\centering
\begin{minipage}[t]{0.23\textwidth}
\centering
\includegraphics[width=4cm]{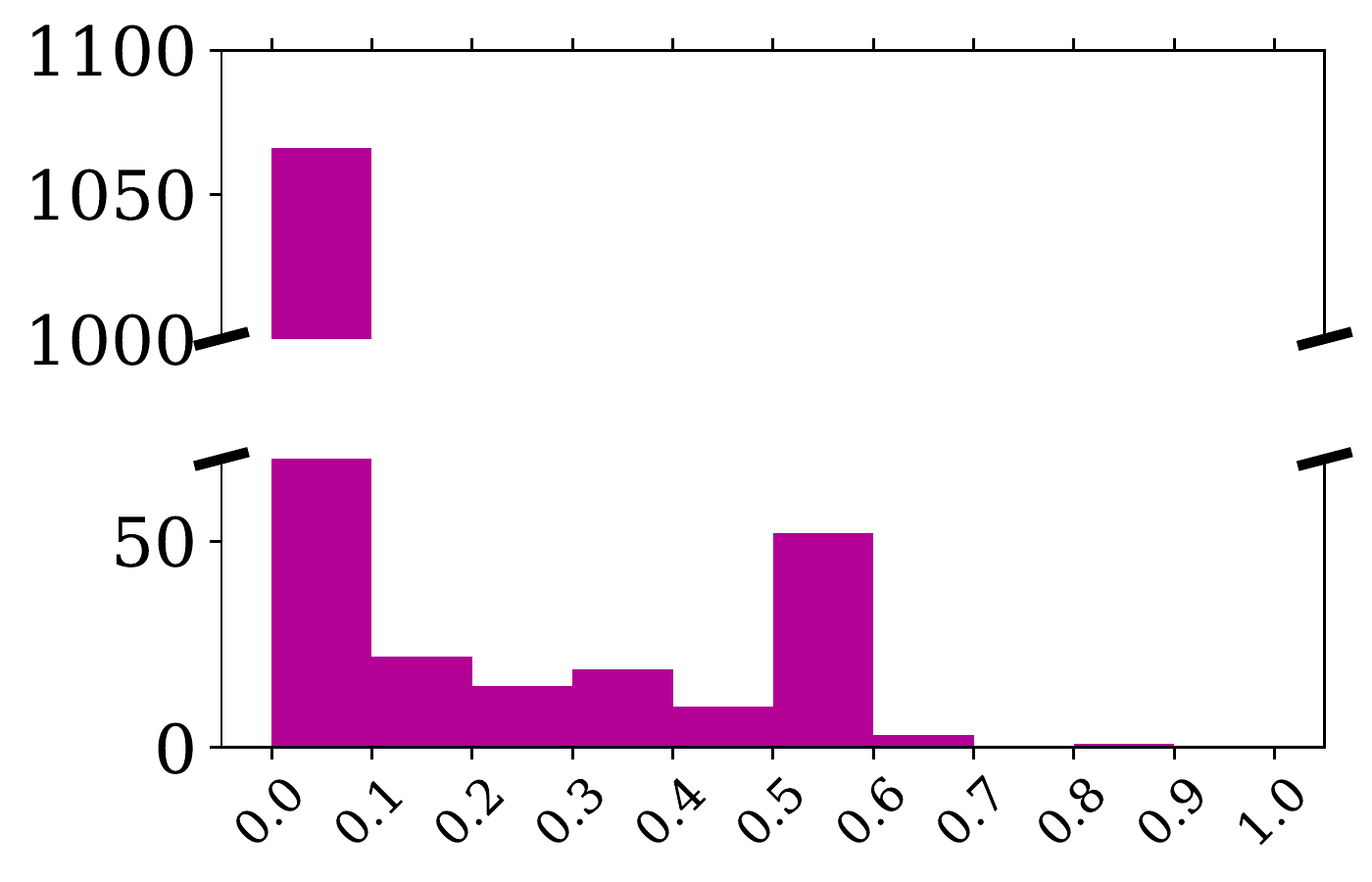}
\caption{Distribution of index $\sigma$.}
\label{fig:sigmadistribution}
\end{minipage}
\begin{minipage}[t]{0.23\textwidth}
\centering
\includegraphics[width=4cm]{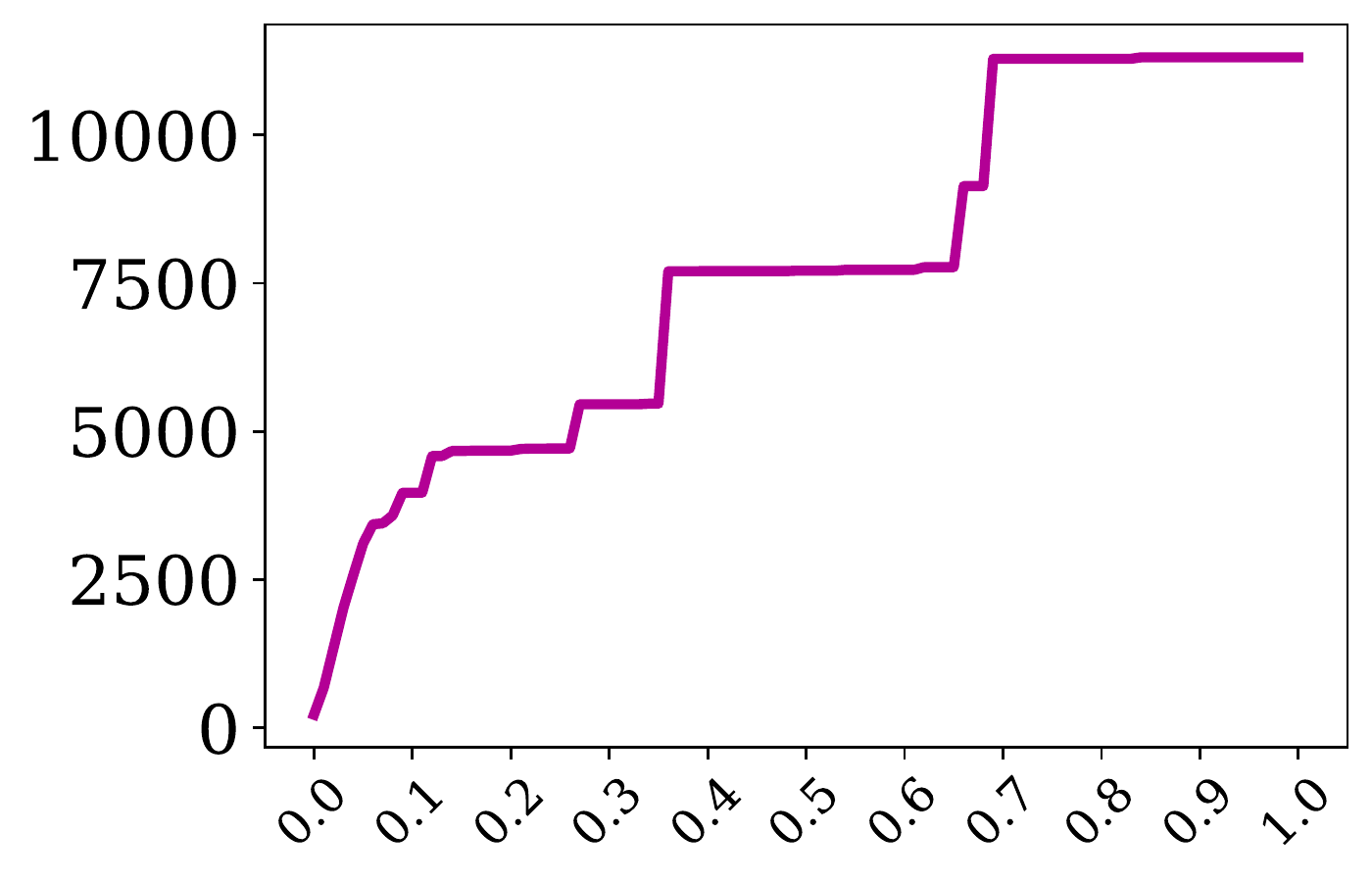}
\caption{Remain amount.}
\label{fig:sigmaremain}
\end{minipage}
\end{figure}

To determine the specific value of the unique index $\sigma$, we empirically investigate the relationships between different values of $\sigma$ and their result. The distribution of $\sigma$ is shown in Figure~\ref{fig:sigmadistribution}, the horizontal axis of which denotes the range of $\sigma$, the vertical axis of which denotes the number of different positions of vendors corresponding to the range of $\sigma$. According to the figure, most of the positions have a few unique tokens (i.e., $\sigma \textless 0.1$). When $\sigma$ is around 0.5, there is a peak, which we think may be the position where a large number of meaningless tokens appear. Moreover, we count the remaining tokens when we adopt a different value of $\sigma$, which is shown in Figure~\ref{fig:sigmaremain}. We find that there is a spike of around 0.3 and 0.7. We speculate that this may be due to the sudden introduction of a large number of meaningless tokens when the values are greater than 0.3 and 0.7, resulting in this phenomenon. Based on the above observations and analysis, we set the $\sigma$ at 0.3.
For each vendor, we separately process the tokens column by column of the corresponding AV labels until $\sigma$ drops below 0.3. We keep scanning the AV labels orderly and reversely, to \textit{fade away} the meaningless tokens, processed result of which is shown in Figure~\ref{fig:runningexample}-3.


\subsection{Vectorization}
In this stage, \sys transforming the filtered tokens into vectors aiming to better describe their relationship in high dimensional space. Prior works\cite{sebastian2016avclass,sebastian2020avclass2} focus on leveraging manually extracted rules to \textit{align} the tokens (i.e., capture the aliases and the similar tokens, and then merge them). The manually-extracted rules lack expansibility to the evolutive trends of malware and labor-intensive. To ensure the \sys's intelligence and expansibility, we turn to unsupervised machine learning to align the tokens, according to the situation of less trustable ground truth equipped and trying to reduce the manpower. 
 

Before vectorizing the tokens, we notice that the very same AV labels are produced by distinct vendors simply because they do rely on a common, third-party component. But we do not merge them because we're not totally dependent on either single vendor. If multiple vendors trust the same component, indicating that the component has higher credibility, it will naturally increase the weight of these AV labels.

The first step to applying a machine learning model is to vectorize the filtered tokens. In our scenario, we need an unsupervised machine learning approach to perform the vectorization procedure. Based on our observation, if two tokens have some connections (e.g., subordination, synonyms, alias, etc.), they will appear together, and the contexts of which are always similar. Therefore, we take the co-occurrence relations as the key factor to model the related words, which are helpful for us to extract and sort the keywords output for users. 
There are two more requirements for the vectorization method during the experiment: parameter insensitive and low computational complexity, which are all intended to allow \sys to self-iterate quickly when it encounters incremental and unseen malicious samples. 

By a thorough investigation, we find GloVe (\textbf{Glo}bal \textbf{Ve}ctors)\cite{pennington2014glove} is a suitable choice in our scenario. GloVe is good at generalizing the co-occurrence relationship in the \textit{global} scope. What's more, according to the algorithm design, GloVe does not use neural networks, and the computational complexity is independent of the data size, which is conducive to the system expansion.
The main procedures of this technique are firstly to count the co-occurrence matrix for each token with a fixed counting window and then to reduce the dimension of the sparse matrix while ensuring the co-occurrence relation between each word. The target function is designed to make related tokens closer in the higher dimension, while irrelated tokens further, shown as follows,

\begin{equation}
J=\sum_{i, j=1}^{N} f\left(X_{i j}\right)\left(T_{i}^{T} \tilde{T}_{j}+b_{i}+\tilde{b}_{j}-\log X_{i j}\right)^{2}
\end{equation}

where $T_{i}^{T}$ denotes the transpose of $i^{th}$ token vector, ${T}_{j}$ denotes the vector of $j^{th}$ token, $b_i$ and $\tilde{b}_{j}$ denote the bias terms, $X_{i j}$ denotes the amount of $j^{th}$ token appearing in the window of $i^{th}$ token, function $f$ denotes the weight function, and $N$ is the size of the token vocabulary.

\subsection{In-Sample Clustering}

In this stage, we regroup the tokens by samples. After being processed by the former steps, we transform the tokens into vectorized representations, trying to map the context-aware relationships to the higher dimensions. \sys aims to output the most related token or token set to the malicious sample. So if the related tokens and the irrelated tokens can be classified into two groups, the output can be produced much more effectively. Fortunately, according to our observation, related tokens are always appearing together (e.g., \textit{Trojan}, \textit{Redirctor} and \textit{Hidelink}), the transformed vectors of which are also closer in the higher dimension. Due to the situation of lacking ground truth (for supervised machine learning), a suitable unsupervised clustering technique can pave the way for the final output.


\paragraph{Token Clustering}

There are several clustering algorithms that adopt different strategies to fit diverse situations. To meet our needs, we have some requirements for the clustering method. \textit{First}, the lower the computational complexity, the better. As the system iterates, more new samples are introduced, and the calculation speed greatly affects the performance of the system. \textit{Second}, the algorithm should free from presetting the number of clusters. Because we can not predict how many clusters will be produced from one sample. 

According to the aforementioned analysis result and some empirical experiments, we leverage Mean Shift\cite{comaniciu2002meanshift} as our clustering algorithm. Mean Shift is a non-parametric feature-space analysis technique, which is always applied in cluster analysis. In other words, the algorithm calculates the offset means of the current point by specifying the radius of a high-dimensional sphere (i.e., parameter \textit{bandwith}) and by continuous iteration, so that it can move to the space containing the most points at the same time. This characteristic quite fits our requirements, and we also hope to find the set of tokens with the highest correlation, where may contain the most accurate descriptive tokens of the malicious samples, as shown in Figure~\ref{fig:runningexample}-4.




\paragraph{Token Correction}
After grouping the tokens according to their relevance, we can sort them according to their frequency and the attributes of the clusters. However, we find that there is no uniform naming convention among the AV vendors, so some of the tokens may imply the same word, but with several different letters, even because vendors come from different countries. There are several reasons why we are considering token correction in the current step. On the one hand, because the computational complexity of pair comparison is $O(n^2)$, it takes a long time to make a comparison in the whole corpus. On the other hand, there may be mis-correction in the algorithm design, and we hope to minimize the side effects.

\begin{figure}[!h]
    \centering
    \includegraphics[width=7cm]{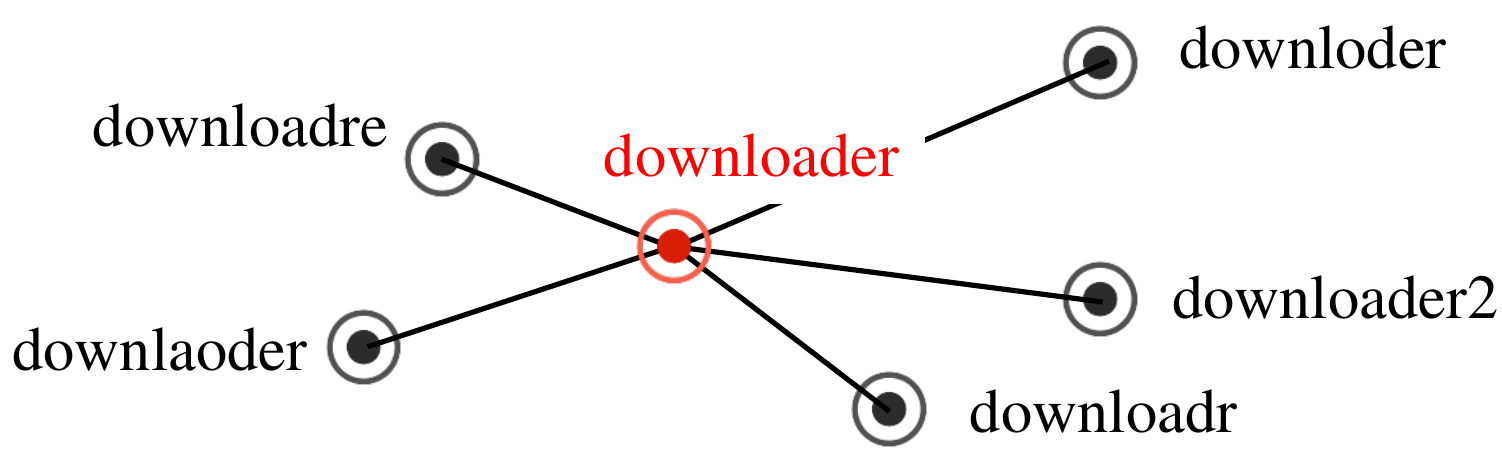}
    \caption{An Example of the Correct Token and Mistaken Tokens.}
    \label{fig:correction}
\end{figure}

As shown in Figure~\ref{fig:correction}, it is a real case from our token corpus. All the tokens imply the same word \textit{downloader}, but only the token colored with {\color{red}{red (downloader)}} is spelt correctly. Because the frequency of the token plays an important role in the final rank, so we may lose the weight of key tokens by mis-spelling. For instance, a malicious sample output the regrouped tokens (in one cluster) as shown in Table~\ref{tab:correction}, ranked by token frequency. Row 1 and 2 denote the tokens and their amount before being corrected, and rows 3 and 4 denote the ones after the procedure. The manually labeled ground truth of this sample is \textit{plankton}, but several vendors mis-spell the words (i.e., plangton), so the key token \textit{plankton} loses some frequency weight, which may lead to the wrong output. Once we correct the words and refresh the rank by frequency (from rank 5 to rank 2), we may have a higher probability of output the right words in the final stage.

\vspace{-8pt}
\begin{table}[!h]
\centering
\caption{Example of token correction.}
\begin{tabular}{c|c|c|c}
\Xhline{1.2pt}
\multicolumn{2}{c|}{{Before Correction}} & \multicolumn{2}{c}{After Correction} \\\Xhline{1.2pt}
Token                        & \#           & Token                    & \#        \\ \hline
andriod                      & 13           & andriod                  & 13        \\
trojan                       & 11           & \textbf{plankton}        & 12        \\
airpush                      & 8            & trojan                   & 11        \\
plangton                     & 6            & airpush                  & 8         \\
\textbf{plankton}            & 6            & adware                   & 4        \\\Xhline{1.2pt}
\end{tabular}

\label{tab:correction}
\end{table}
\vspace{-8pt}

To complete token correction without affecting normal tokens, two prerequisites are given. Firstly, the correction task is only limited to the same cluster within the sample, minimizing the computational complexity (the complexity of pair comparison is $O(n^2)$). Secondly, We raise an error correction threshold $\delta$ to detect the tokens that need to be corrected, and the token will be corrected only when it reaches this threshold, which is defined as follows,

\begin{align}
    \delta = \frac{\mathsf{Edit}(\mathsf{T}_i, \mathsf{T}_j)-\mathsf{LenDiff}(\mathsf{T}_i, \mathsf{T}_j)}{\mathsf{MaxLen}(\mathsf{T}_i,\mathsf{T}_j)}
\end{align}

where $\mathsf{Edit}(\cdot, \cdot)$ denotes the Edit Distance (minimum number of edits for string conversion) between 2 tokens (i.e., $\mathsf{T}_i$ and $\mathsf{T}_j$), $\mathsf{LenDiff}(\cdot, \cdot)$ denotes the length difference, and $\mathsf{MaxLen}(\cdot, \cdot)$ denotes the max length among them. The reason for setting the threshold in this way is that, not only can it correct a few different token pairs (e.g., \textit{downloader} and \textit{downloadre}), but also it can also find the abbreviations of token pairs (e.g., \textit{gen} and \textit{generic}). When the threshold $\delta$ is below 0.3, we think there are candidates in the token pair that need to be corrected, and if a \textit{standard word} (according to Wiki) does not exist in them until a \textit{standard word} associated with them is found. The result is shown in Figure~\ref{fig:runningexample}-5.

\subsection{Keywords Output}

To enhance the readability of the output, we try to rerank the tokens, which also benefits customizing the amount of output. Besides the clustering relations and token frequency, which can be leveraged as ranking criteria, more global information should also be taken into consideration. Thanks to the technique TF-IDF (Term Frequency–Inverse Document Frequency)\cite{jing2002improvedtfidf}, which is a common weighting technique for information retrieval and data mining, it is a statistical method used to assess the importance of a word to one of the documents in a set of documents or corpus. The importance of a word increases proportionally with the frequency of its occurrence in the document but decreases inversely with the frequency of its occurrence in the corpus. In other words, this method will pick out the tokens that are relevant and important, which will be ranked higher. The method is calculated as,

\begin{equation}
{T F_{T_i, L(mal_j)}=\frac{\operatorname{count}(T_i)}{\left|L(mal_j)\right|}}
\end{equation}

$TF_{T_i, L(mal_j)}$ (i.e., abbreviation of \textit{\textbf{T}erm \textbf{F}requency}) denotes the frequency of token $T_i$ in the malicious sample $L(mal_j)$, $count(T_i)$ denotes the frequency amount of token $T_i$, and $\left|L(mal_j)\right|$ denotes the amount of all the tokens in the malicious sample $L(mal_j)$.

\begin{equation}
IDF_{T_i}=\log \frac{N}{\sum_{j=1}^{N} I\left(T_i, L(mal_j)\right)}
\end{equation}

$IDF_{T_i}$ (i.e., abbreviation of \textit{\textbf{I}nverse \textbf{D}ocument \textbf{F}requency}) reflects the universality of token $T_i$ in the corpus, $N$ denotes the amount of malwares, $I\left(T_i, L(mal_j)\right)$ denotes whether the malware $L(mal_j)$ contains the token $T_i$ (1 for positive and 0 for negative). However, if the token $T_i$ does not appear in all the malwares (e.g., when the dataset changes), then the denominator in the fomula equals to 0. So we need to smooth it, which is shown as below,

\begin{equation}
IDF_{T_i}=\log \frac{N}{1+\sum_{j=1}^{N} I\left(T_i, L(mal_j)\right)}
\end{equation}

with the equations above on hold, we can calculate the $TF-IDF$ value as follows,

\begin{equation}
TF-IDF_{T_i, L(mal_j)}=TF_{T_i, L(mal_j)} * IDF_{T_i}
\end{equation}

According to the definition and formulas, when the frequency of a token in the malicious sample is higher, and the rarity is higher (i.e., the universality is lower), its TF-IDF value is higher. The method greatly fits our need which gives consideration to token frequency and rarity, filters some common words and preserves vital words that can provide more information. And we can calculate the TF-IDF value of each token in each malicious sample.

By utilizing the clustering relations, token frequency, and TF-IDF value, the tokens can be reranked in each malicious sample by several steps, which is shown in the following pseudocode. Firstly, we need to find the \textit{best cluster} as the candidate cluster, which contains the token with the highest TF-IDF value. According to our observation, the \textit{best cluster} always comprises the tokens highly related to each other and to the malicious samples. Next, by judging whether the amount of tokens in the best cluster is sufficient, we will treat different situations separately. If the amount is sufficient, the first N tokens in the best cluster are added to the final result. And if the token has the second-highest TF-IDF value, which also indicates a relatively high correlation, we will replace it with the $N^{th}$ token in the result.  
Otherwise, if the amount is insufficient, all the tokens in the best cluster are added to the result, the rest of which will be filled by the tokens according to the TF-IDF value. Finally, the result is presented to users in reranked tokens as shown in Figure~\ref{fig:runningexample}-6.

\begin{algorithm}  
\scriptsize
\caption{Token rerank and output}  
\LinesNumbered  
\KwIn{\textbf{Clusters:} Token clusters, as a 2D array. Each array, Clusters[ClusterID] denotes a token cluster, as an array of tokens, where the rank of the tokens indicates the token frequency.\newline
    \textbf{TFIDF:} Array of tokens, ranked by TFIDF  \newline
    \textbf{TopN:} Amount of output tokens   
}
\KwOut{\textbf{Result:} Array of tokens, sorted by importance}  
\textbf{Initialize:} Result = $\emptyset$

\tcc{Find the ID of the Best Cluster}
\For{ClusterID from 1 to Clusters.Length}{
    \If{TFIDF[1] in Clusters[ClusterID]}{
    BestClusterID $\leftarrow$ ClusterID
    }
}

\tcc{The Amount of Tokens in the Best Cluster is Sufficient}
\If{Clusters[BestClusterID].Length $\ge$ TopN}{
    Result $\leftarrow$ Clusters[BestClusterID][1:Top(N)]
}
\If{TFIDF[2] not in Clusters[BestClusterID]}{
    Result[TopN] $\leftarrow$ TFIDF[2]
}

\tcc{The Number of Tokens in the Best Cluster is Insufficient}

Result $\leftarrow$ Clusters[BestClusterID]

\For{Index=1; Result.Length $\textless$ TopN \newline
and Index $\textless$ TFIDF.Length; Index+=1}{
\If{TFIDF[Index] not in Result}{
Result $\leftarrow$ Result+TFIDF[Index]
}
}
\end{algorithm}

\section{Evaluation}

In this section, we first describe the implementation in detail. Then, the dataset we use to evaluate \sys is introduced,  followed by the comparison of tagging accuracy with prior works \AVCLASS and \AVCLASSTWO. Finally, we measure \sys's robustness in terms of dataset size, file type, and sample detected time. 


\subsection{Implementation}
In this section, we present the implementation of our system, including the specific experimental details and parameters.

Firstly we retrieve data from VirusTotal. Once the retrieval finishes, we analyze all the data to filter the tokens, removing potentially meaningless tokens, using $\sigma$=0.3. Afterward, we train a GloVe model, where the window size is 40, the vector length is 32, and the training epoch is 100. Then, we use Mean Shift, where bandwidth is 2 and training epoch is 100, and TF-IDF to cluster tokens inside each malicious sample. Finally, we correct tokens inside each cluster with $\delta$=0.3.

The Python implementation uses the following libraries: \textit{glove}\cite{glove} and \textit{gensim}\cite{gensim} for vectorization, \textit{scikit-learn}\cite{scikit} for TF-IDF, \textit{editdistance}\cite{edit} for words correction, \textit{tabulate}\cite{tabulate} for visualization. The evaluation part uses GNU Parallel\cite{gnuparallel} to speed up.  

All our experiments are conducted on a PC with 16 GB memory, 1 Intel i7-7700k CPU (4.2 GHz).




\subsection{Dataset}
\label{sec:dataset}




We evaluate \sys on the datasets shown in Table~\ref{tab:dataset}. Drebin\cite{arp2014drebin} and Malheur\cite{rieck2011automaticmalheur} are manually collected and labeled malware datasets, which are also used in prior works. These two datasets are well-known and also widely used by several related works. 
Please note that our \sys is free from domain knowledge, and it can adapt to newly come malicious samples with fine-tuned models.

We also get a Superset by combining the two datasets. Column 2 to 5 separately denotes the platform, the number of malicious samples, the time range of sample collection, and whether the dataset has manually labeled ground truth.  

\vspace{-6pt}
\begin{table}[!h]
\scriptsize
    \centering
    \caption{Datasets used in the evaluation.}
    \begin{tabular}{l|c|c|c|c}
    \Xhline{1.2pt}
    \textbf{Dataset}    &  \textbf{\# of Virus Type} & \textbf{Amount} & \textbf{Time Range} & \textbf{with G.T.} \\\Xhline{1.2pt}
    Malheur\cite{rieck2011automaticmalheur}  & 24 & 3,086 & 08/2009 - 08/2009 & \cmark\\
    Drebin\cite{arp2014drebin}  & 178 & 5,511 & 08/2010 - 10/2012 & \cmark\\
    \Xhline{0.5pt}
    \textit{Superset}     &  202 & 8,597 & 08/2009 - 10/2012 & \cmark \\
    \Xhline{1.2pt}
    \end{tabular}
    
    \label{tab:dataset}
\end{table}
\vspace{-8pt}

We present the statistical results on file type as follows to show the dataset more clearly.






\begin{table}[!h]
\parbox{.48\linewidth}{
\scriptsize
\centering
\caption{Malheur (Type)}
\begin{tabular}{c|c}
\Xhline{1.2pt}
File Format & Amount \\\Xhline{1.2pt}
Win32 EXE & 2945\\
DOS EXE & 139\\
unknow & 2\\\Xhline{1.2pt}
\end{tabular}

\label{tab:maltype}

}
\hfill
\parbox{.48\linewidth}{
\scriptsize
\centering
\caption{Drebin (Type)}
\begin{tabular}{c|c}
\Xhline{1.2pt}
File Type & Amount \\\Xhline{1.2pt}
Android & 5503\\
ELF & 1\\
GZIP & 1\\
ZIP &1\\
unknown & 5\\\Xhline{1.2pt}
\end{tabular}

\label{tab:dretype}}
\end{table}

Table~\ref{tab:maltype} and~\ref{tab:dretype} present the distribution of file formats (i.e., a standard way that information is encoded for storage in a computer file) in our datasets. Malheur mainly consists of windows executables (95.4\%), with several DOS executables (4.5\%) and 2 samples that cannot be classified by VirusTotal. Meanwhile, Android accounts for 99\% of dataset Drebin, with a few ELF, GZIP, ZIP, and 5 undetectable samples. They contain malicious samples from 2 mainstream mobile and PC platforms, based on which we can better demonstrate that \sys is available for the commonly used platforms.


\subsection{Overall Accuracy}

Our experiments do not introduce external data without clarity to augment the dataset described in Section~\ref{sec:dataset}, which can show that our \sys is not dependent on a large amount of available data and that even a small amount of incremental data can yield substantial results. 


\begin{table}[!h]
\centering
\scriptsize
\caption{Tagging Accuracy of Malheur and Drebin.}
\begin{tabular}{c|ccc|ccc}
\Xhline{1.2pt}
\multirow{2}{*}{\textbf{TopN}} & \multicolumn{3}{c|}{\textbf{Malheur}}  & \multicolumn{3}{c}{\textbf{Drebin}}   \\
                       & \sys & AV\tiny{CLASS} & AV\tiny{CLASS2} & \sys & AV\tiny{CLASS} & AV\tiny{CLASS2} \\\Xhline{1.2pt}
1                      & 0.744   & 0.811   & 0.376    & 0.866   & 0.899   & 0.039    \\
2                      & 0.859   & 0.811   & 0.558    & 0.914   & 0.899   & 0.897    \\
3                      & 0.896   & 0.811   & 0.570    & 0.973   & 0.899   & 0.933    \\
4                      & 0.919   & 0.811   & 0.721    & 0.979   & 0.899   & 0.940    \\
5                      & 0.926   & 0.811   & 0.782    & 0.980   & 0.899   & 0.942    \\
6                      & 0.946   & 0.811   & 0.799    & 0.981   & 0.899   & 0.942    \\
7                      & 0.947   & 0.811   & 0.799    & 0.982   & 0.899   & 0.942    \\
8                      & 0.948   & 0.811   & 0.800    & 0.983   & 0.899   & 0.942    \\
9                      & 0.948   & 0.811   & 0.802    & 0.985   & 0.899   & 0.942    \\
10                     & 0.950   & 0.811   & 0.830    & 0.985   & 0.899   & 0.942   \\\Xhline{1.2pt}
\end{tabular}

\label{tab:tagcomparemaldrebin}
\end{table}

Table~\ref{tab:tagcomparemaldrebin} presents the evaluation result on dataset Malheur and Drebin, by comparing \sys with prior works \AVCLASS and \AVCLASSTWO. Row 3 to 12 denote the tagging accuracy with $TopN$ ranging from 1 to 10 (i.e., if the manually labeled ground truth is the same as either token in the result of the concatenation of them, we take it as the positive case). Since \AVCLASS produces only one token for a single malicious sample, its accuracy will remain the same no matter how $TopN$ changes. When we choose Top1, \AVCLASS performs better on both datasets than others, because it gives a very clear and condensed classification result for a malicious sample, with the help of rich expert knowledge. Once we choose Top2, \sys goes beyond other tools. As we expand the chosen range, the tagging accuracy of \sys continues to rise. The reason for that \sys cannot precisely hit the ground truth within Top1 is that \sys sometimes place some generic tokens at first, such as \textit{Android}, \textit{JS}, etc. In our system design, we do not remove the tokens describing the file format, which is a benefit for users to more comprehensively understand the malicious samples. So the tokens describing categories of behaviors may be placed to the back.
But when the chosen scope is expanded, they come into the output token list. 


\begin{figure}[!h]
\centering
\begin{minipage}[t]{0.23\textwidth}
\centering
\includegraphics[width=4cm]{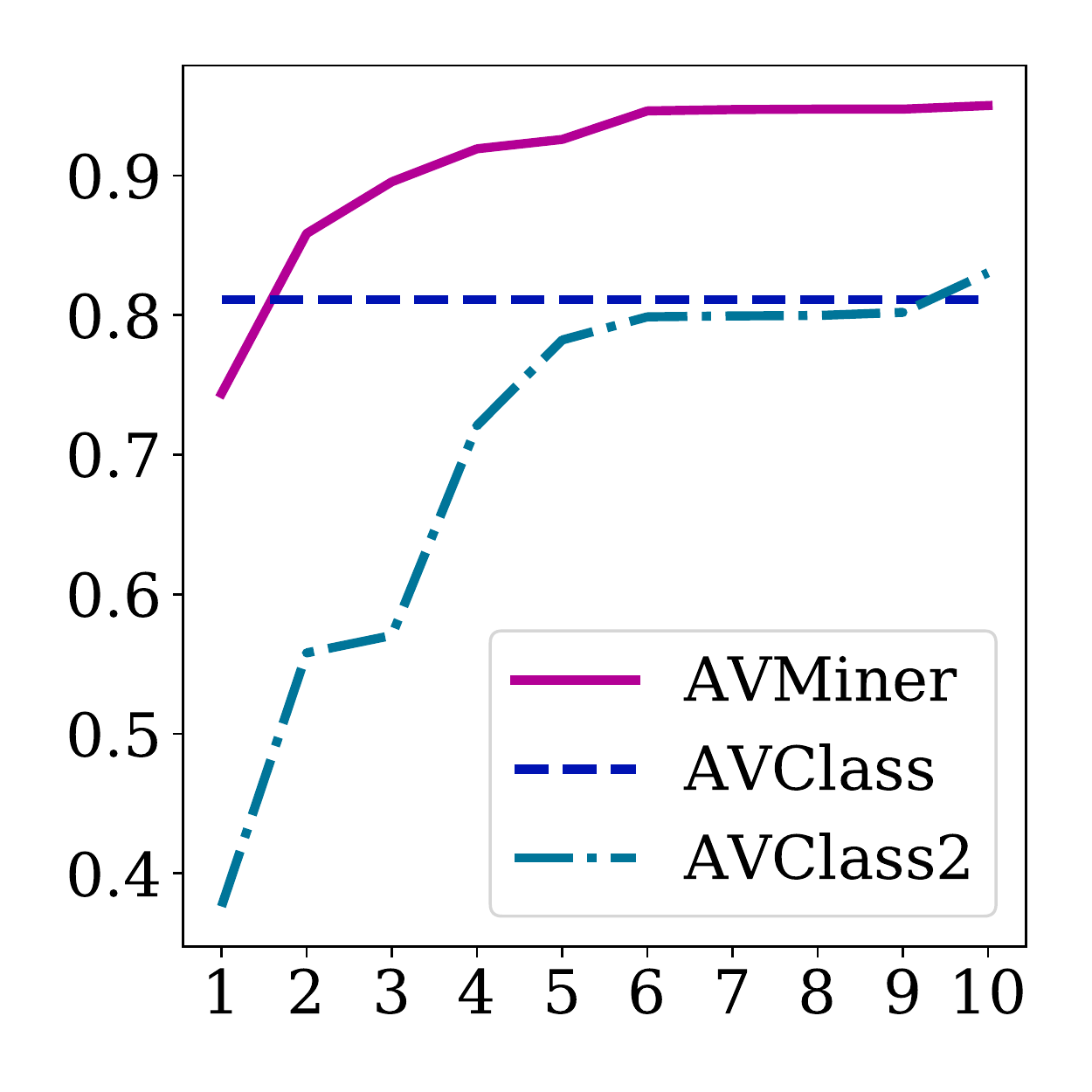}
\caption{\footnotesize{Malheur ACC. in Top1-10}}
\label{fig_22}
\end{minipage}
\begin{minipage}[t]{0.23\textwidth}
\centering
\includegraphics[width=4cm]{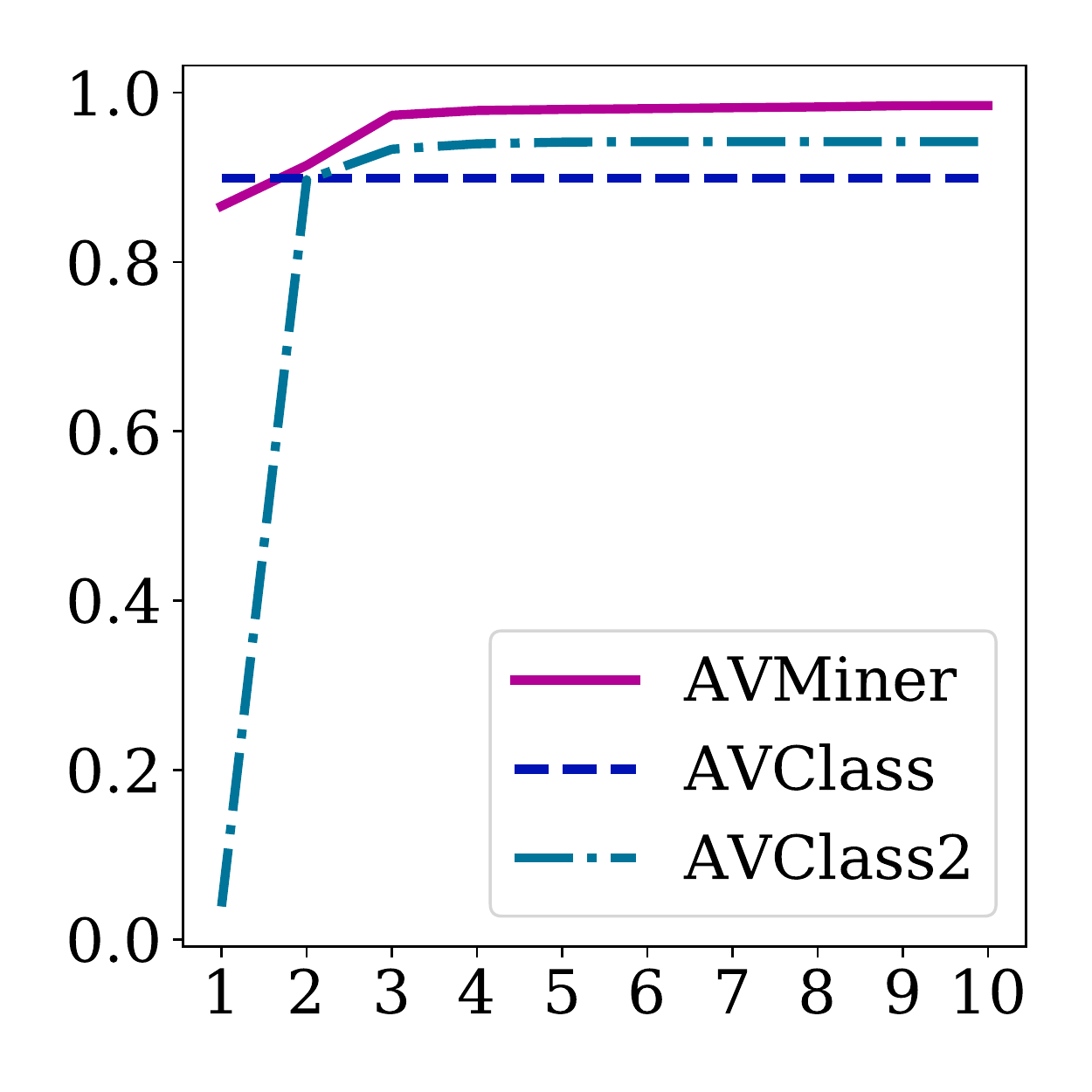}
\caption{\footnotesize{Drebin ACC. in Top1-10}}
\label{fig_23}
\end{minipage}
\end{figure}

To show the changing trend of accuracy with the expansion of the selection range (i.e., $TopN$), we present the results in Figure~\ref{fig_22} and ~\ref{fig_23}. In the Malheur dataset, \AVCLASS performs poorly as the baseline, while \AVCLASSTWO surpasses \AVCLASS until $TopN$ expanded to 10. Although \sys is on a rising trend, it stabilizes at Top6. Produce such an amount of tokens for the users is quite reasonable. According to Figure~\ref{fig_23}, all the tools achieve good results in dataset Drebin, and when the result contains more than 3 tokens, the accuracy is very stable ever since.

To further explore the different performances (accuracy and trend) on the different datasets, we try to deeply analyze the result and find the root cause. Intuitively, we take the diversity of virus type, the amount of the dataset, and the time range of collecting the samples.
We firstly investigate the tagging accuracy on different virus types of ground truth. In Malheur, there are 24 types of virus, which are relatively evenly distributed, with \textit{allaple}, \textit{podnuha}, and \textit{rotator} having the most samples, all with 300. While in Drebin, the taxonomy is more detailed. There are 178 virus types in all, type \textit{fakeinstall} has the most amount of samples (907), and some virus types have only one sample (e.g., \textit{acnetdoor}, \textit{cellshark}).


We investigate the accuracy of each virus type. Even Drebin contains a more diverse virus type of malicious samples, the majority of the dataset is made up of a few kinds of virus types, e.g., \textit{Fakeinstaller}, \textit{DroidKungfu}, and \textit{Plankton}. Most of the AV vendors can effectively label these categories, so it paves the way for the extraction. However, sometimes most of the AV vendors cannot label the malicious samples, so it blocks us from label extraction.
As for Malheur, one of the majority labels, Spygames, which takes up 4.5\%, fail to be distinguished by almost all of the AV vendors, only except for AntiVir. What's worse, the output of AntiVir for this type of malicious samples is "TR/Spy.Games.A", causing all 3 tools' tokenization strategy to split it as "Spy", "Games", and "Spy" and "Games" do not appear in the \textit{Best Cluster}, which leads to the failure. As for \AVCLASS and \AVCLASSTWO, they rank the labels only on their frequencies, causing ignorance. 
On the other hand, we could extract some categories (e.g., \textit{rbot}, \textit{zhelatin}) with 100\% accuracy but \AVCLASS and \AVCLASSTWO miss all of the samples. We speculate that the expert knowledge-based method sometimes leads to failure since it cannot cover all the situations. But we choose to benefit from all the vendors, which equips \sys with more expansibility and adaptiveness.

\begin{table}[!h]
\centering
\scriptsize
\caption{Tagging Accuracy Comparison of the Superset}
\begin{tabular}{c|ccc}
\Xhline{1.2pt}
\multirow{2}{*}{\textbf{Top-N}} & \multicolumn{3}{c}{\textbf{Superset}} \\
                      & \sys & AV\tiny{CLASS} & AV\tiny{CLASS2} \\\Xhline{1.2pt}
1                      & 0.800   & 0.867   & 0.161    \\
2                      & 0.910   & 0.867   & 0.774    \\
3                      & 0.935   & 0.867   & 0.802    \\
4                      & 0.966   & 0.867   & 0.860    \\
5                      & 0.979   & 0.867   & 0.884    \\
6                      & 0.981   & 0.867   & 0.890    \\
7                      & 0.989   & 0.867   & 0.890    \\
8                      & 0.989   & 0.867   & 0.891    \\
9                      & 0.990   & 0.867   & 0.891    \\
10                     & 0.991   & 0.867   & 0.902   \\\Xhline{1.2pt}
\end{tabular}

\label{tab:supersetacc}
\end{table}


Since \sys employs unsupervised machine learning, we suspect that the performance may rely on the distribution and amount of malicious samples. We apply the same experiment on the Superset, and the result is shown in Table~\ref{tab:supersetacc}. The result shows that the performance on superset is even better than the weighted average of the subsets. Parts of the datasets may gain from each other (i.e., context relations), improving the performance of previously underperforming parts. If so, it indicates that the more related samples introduced, the better the performance will be.

According to the result, the accuracy of the previously mentioned sample "Spygames" increases a lot, up to 50\%. The reason is that the token "Spy" appears in Drebin with richer context information, i.e., "Spybubble", "Spymob", "Spyphone", etc. Therefore, this has a positive effect on establishing the relation between "Spygames" and other tokens, which introduces the token "Spy" and "Games" into the \textit{Best Cluster}.

\subsection{Amount Sensitive}

The previous results indicate that the performance of \sys may be sensitive to the number of malicious samples. With a larger amount, the diverse token context can provide much more contextual information, which will enhance the correlations between the related tokens. It will help \sys more easily to capture the token relations in the real world. 

\vspace{-6pt}
\begin{figure}[!h]
    \centering
    \includegraphics[width=7cm]{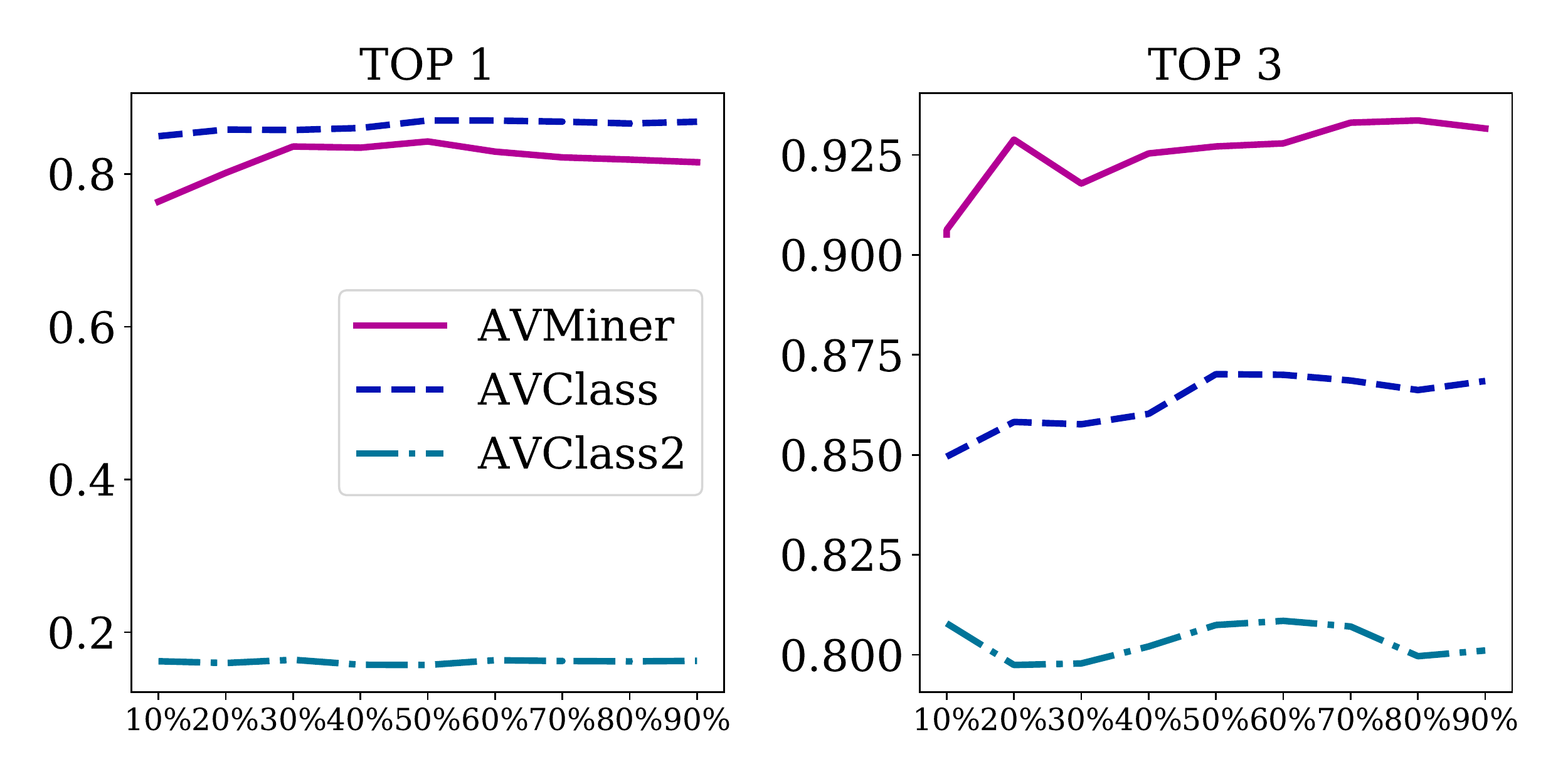}
    \caption{Subset Evaluation result (Top1 and Top3) Ranging from 10\% to 90\% of Superset}
    \label{fig:supersetsubset}
\end{figure}
\vspace{-8pt}

To validate our thought, we evaluate \sys on datasets of different sizes, ranging from 10\% of the superset to 90\% of the superset, comparing to the result by using the whole superset. Our tool aims to provide users precise information with limited words. According to previous experiments, the accuracy becomes stable after TopN is larger than 3. So we choose Top1 and Top3 as representatives in the following experiments.

As shown in Figure~\ref{fig:supersetsubset}, the subfigure on the left denotes the result of Top1, and the right side denotes the result of Top3. To ensure the fairness of sampling, we conduct these experiments 10 times each. We plot an area, and its upper bound and lower bound refer to the result range of the experiment of each subset. The line inside the area denotes the average result. 
We find that the size of the dataset indeed has some influence. The accuracy of \sys slightly drops when the dataset is relatively small. When the size of the dataset reaches 2,400 (about 30\% of Superset), its accuracy doesn't differ much from the whole superset. Additionally, as the amount of data increases, the accuracy grows larger, and the fluctuation of accuracy also reduces. 
According to the result, the larger the data set, the richer the samples, the closer their distribution to the real world, and the more stable the results.
Compared to \AVCLASS and \AVCLASSTWO, we regard the different data distribution as the root cause of the accuracy fluctuation. 

To conclude, the experiment of negative sampling the Superset shows that the number of samples has some impact on the performance. Since the diversity of samples decreases due to the negative sampling, with the faded contextual information, some correlations between the tokens may become alienated, resulting in the result shown above.


\vspace{-12pt}
\begin{figure}[!h]
    \centering
    \includegraphics[width=7cm]{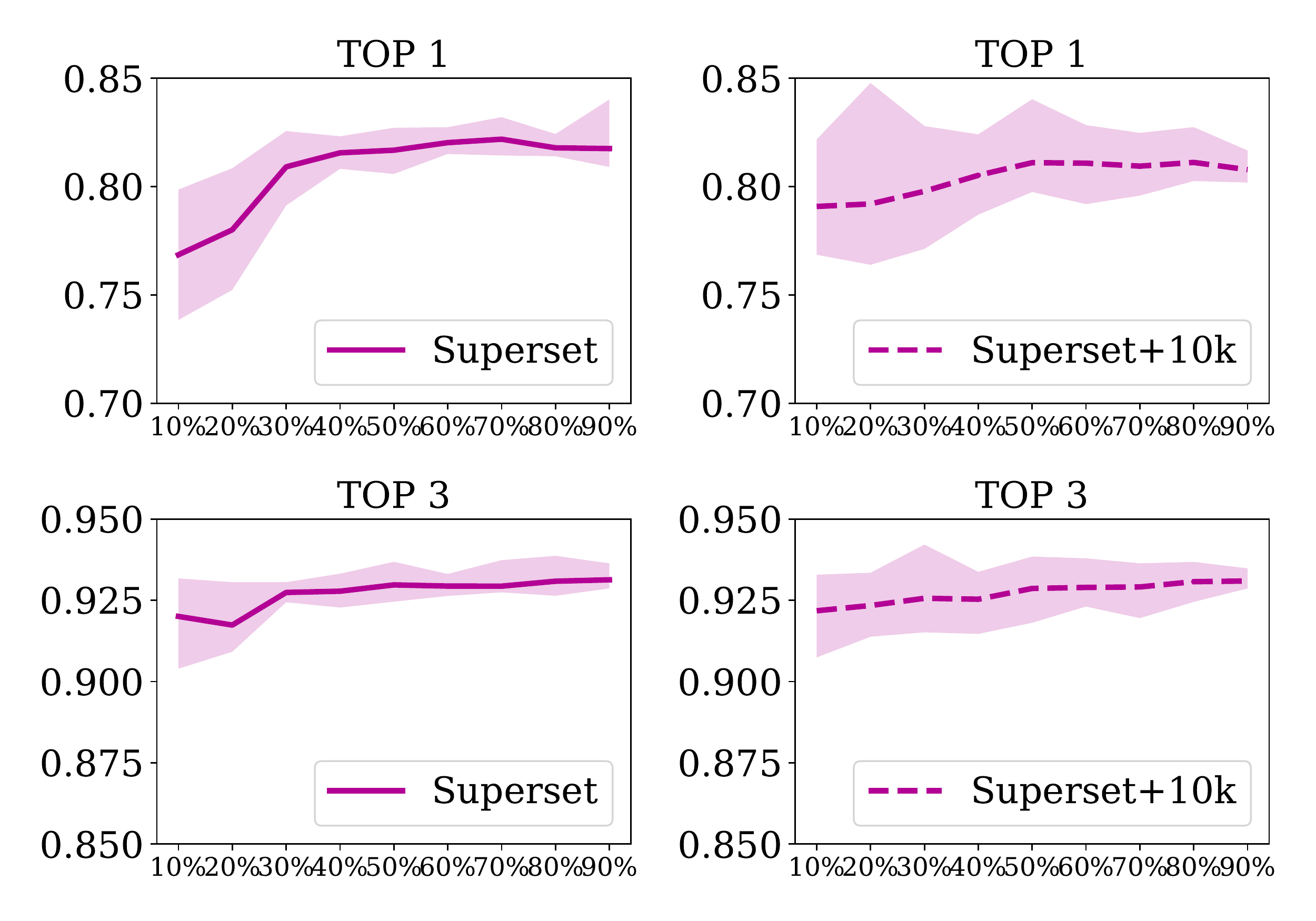}
    \caption{Result comparison of sampling data and with external malicious samples on Top1 and Top3.}
    \label{fig:supersetAUG}
\end{figure}
\vspace{-12pt}

The subsets collected by sampling from the same dataset have a highly similar distribution with its superset. So the result is unpredictable when external data comes into the dataset and breaks the original distribution. To verify the robustness of \sys, we conduct an experiment on the subsets downsampling from the Superset, ranging from 10\% to 90\%, adding 10,000 malicious samples randomly collected from the real world (with no manually labeled ground truth). Due to the data source, we can only evaluate the result on the samples with ground truth, and we also repeat the experiments 10 times for the sampling fairness.

As shown in the Figure~\ref{fig:supersetAUG},
the left side of the figure denotes the Top1 and Top3 results of all subsets, and the right side of the figure denotes the Top1 and Top3 results of the subsets together with 10,000 randomly selected samples.
When random external samples come in, it does lose some accuracy in Top1, but the overall trend is pretty close. According to Top3, the final average accuracy is almost identical. What's more, whether Top1 or Top3, when external data is introduced, the experimental results will be more volatile than the original data (have a wider range of colored area). The experimental results confirm that different distributed data may lead to the loss of accuracy, but the overall stability of the system is still very strong.

\subsection{File Format Sensitive}

To figure out what kind of external uncertainty may influence the robustness of \sys, we firstly control the variable of the type of malicious samples.

\begin{figure}[!h]
    \centering
    \includegraphics[width=5.5cm]{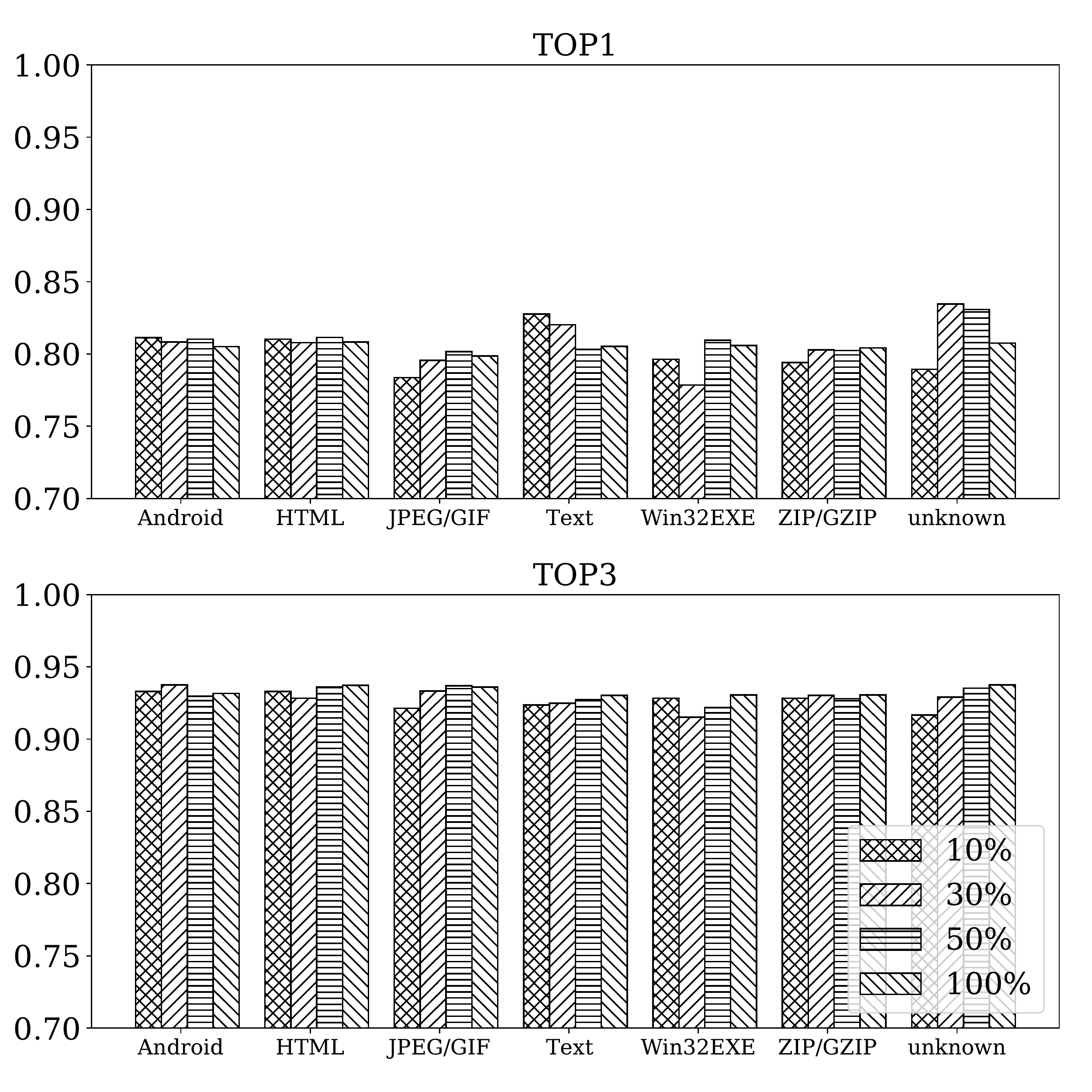}
    \caption{Evaluation Result of Subset with Different Virus Type.}
    \label{fig:supersettype}
\end{figure}

We evaluate \sys on the datasets containing 10\%, 30\%, 50\%, and 100\% of the Superset, all of which are introduced with 10,000 different types of malicious samples from the real world, the result of which is shown in Figure~\ref{fig:supersettype}.
According to the result, it presents the Top1 accuracy only of the subsets on the left and the Top3 on the right. Due to the introduction of external samples, \sys is relatively unstable in the Top1 results verified in the previous experiments. Fluctuations of accuracy are no more than 5\% in this set of data under our expectations. In contrast, the results of Top3 are more stable, with fluctuations even less than 2\%, which convinces us that under different sampling conditions, our methods will not lead to large fluctuations in results due to different types of data introduced.

\vspace{-12pt}
\vspace{-12pt}

\subsection{Time Range Sensitive}

In this part, we control the variable of the time and conduct a similar experiment as the previous part to test whether the robustness of \sys will be affected by the samples with different time distribution. To clarify, the time we mentioned here represents the time that the malicious samples are first detected by VirusTotal, rather than the latest update time.

\begin{figure}[!h]
    \centering
    \includegraphics[width=8.5cm]{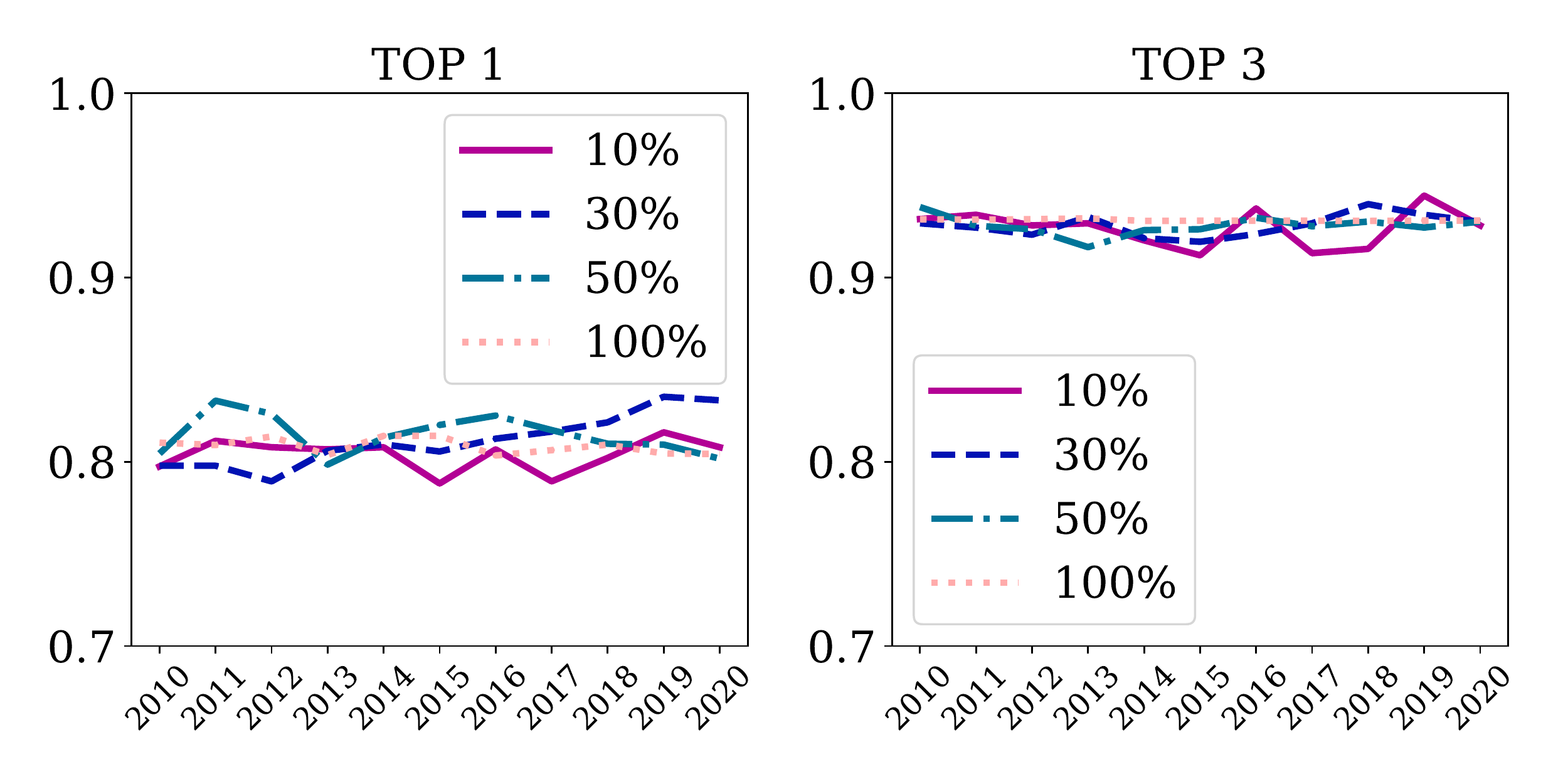}
    \caption{Evaluation Result of Subset with Different Time Range.}
    \label{fig:supersettime}
\end{figure}

As shown in Figure~\ref{fig:supersettime}, it presents the result by evaluating \sys on the datasets containing 10\%, 30\%, 50\%, and 100\% of the Superset, all of which are introduced with 10,000 malicious samples with different detected time, ranging from the year 2010 to 2020.

According to the result, Top1 shows the fluctuations of accuracy, no more than 5\%, similar to the result in type-sensitive experiment. But the Top3 shows that the datasets with newly introduced samples, ranging from 2011 to 2014, which is closer to the year of the collection time of the Superset, have a quite stable performance. When the introduced samples are firstly detected later than 2015, the result shows obvious fluctuations (about 5\%) on Top3 result, while under the same experimental settings, the fluctuation does not exceed 2\% without the influence of external data. 

To find out the root cause, we investigate how the sampled malicious samples change over time. On the one hand, over the past decade, we have become more and more dependent on mobile devices (i.e., depend more on operating systems such as Android, iOS, etc.), and the increasing attention to them is bound to increment security risks. So the platform of malicious samples changes a lot, but it may not the primary cause. More importantly, with the evolution of the arms race in the field of software security, the malicious behavior of the virus has changed a lot. Therefore, in our method, the context of tokens changes, which also leads to the occurrence of fluctuations.

\section{Discussion}

\subsection{Influence of Inconsistent Distribution}

According to the evaluation section, \sys is verified that it better performs than the previous works no matter how the amount and types change, indicating the advancement and robustness. But when we introduce external samples, \sys has some fluctuations in the performance, which is not only due to the sampling strategy. We find that the main factor causing the fluctuation is the time dimension. Based on the current system design, we propose a time grouping based scheme to solve this problem. By grouping the samples at different time range (by year or month) and separately processing them in different groups, the final results are reunion from different groups. Because we do not manually label the most recent malicious samples, the thought is hard to verify. This part will be one of our future works.

\subsection{Performance on the Large Scale Samples}

\sys can give full play to its value of saving manpower when labeling a large number of malicious samples. To verify the effectiveness, the following experiments are added. We collect about 2 million more malicious samples from the real world and try to label them by extracting the AV labels produced from VirusTotal.  
\sys can produce a token set for all the malicious samples as long as VirusTotal returns the AV label of them. However, the labeling strategy of \AVCLASS can only cover 79.1\% of malicious samples, and the rest of them are tagged with "Singleton" indicating no result. Coincidentally, \AVCLASSTWO can produce the label for about 89.4\% of samples, the rest of which are tagged with nothing.
Since the number of malicious samples from each year is not even, we only analyze the proportion of the data that cannot be tagged in each year. According to our analysis, the newer the samples, the higher the probability of their tagging failure. In 2020, about 40\% of malicious samples cannot be tagged by \AVCLASS, while 28.8\% of malicious samples cannot be tagged by \AVCLASSTWO. This further proves that such labeling systems need to be highly adaptable and scalable.

\subsection{Time Overhead}

Intuitively, \sys consumes more time than previous works. The procedures of vectorization and clustering take up most of the time for the multiple iterations. According to our evaluation, \sys can process the output for about 40 samples per second. Even though the rule-based methods are dozens of times faster, We think this is fast enough to meet our daily needs. Fortunately, all the procedures are almost linear relations between the number of malicious samples. It will be our future work to speed up and parallel some procedures.

\section{Related Work}


AV labels benefit the software security society for many applications, such as building malware dataset, malware detection and clustering\cite{bayer2009scalable,jang2011bitshred,kotzias2015certified,perdisci2008mcboost,perdisci2010behavioral}. Such as, Drebin\cite{arp2014drebin} and Malheur\cite{rieck2011automaticmalheur} separately set up datasets on different platforms, and propose tools to detect them. However, AV labels are not always stable. Zhu et al.\cite{zhu2020measuring} survey on a large amount of method for malware labeling and analyze the problem of label inconsistency. 
They trace the AV labels of 14K samples for a year and conduct a thorough analysis indicating that there is influence between vendors on producing AV labels. 
AV-Meter evaluates the performance of AV vendors on different datasets, and the presented results block researchers from using the AV labels straightforwardly.

\textbf{Malware Labeling.} Facing the challenge above, \sys mainly focuses on mining the meaningful tokens from inconsistent AV labels produced by the third-party platform VirusTotal. We take it as an alternative to generate malware labels for the malicious samples, so do \AVCLASS\cite{sebastian2016avclass} and \AVCLASSTWO\cite{sebastian2020avclass2}. The procedures of three methods are summarized and abstracted into three modules: pre-processing, alignment, and output.
First, all of the methods process the AV labels into tokens minimizing the noise. Second, aligning the tokens in terms of format and semantics (merging the similar tokens, etc.) ready for output. Finally, all the methods will output the result with a different ranking strategy for the users. The former works mainly leverage expert knowledge to set up the taxonomy, but \sys substitutes the procedures with intelligent ones. This design indeed causes more processing time, but it lights a way to label the malicious samples with less labor efficiently. 
\section{Conclusion}


In this work, we propose \sys, an expansible AV label mining method. \sys takes as input a malicious sample with its corresponding AV labels, then pre-process, vectorize, group, and output the keywords. Without any expert knowledge, we are fully benefiting from the co-occurrence relations between tokens. According to the evaluation, \sys has better performance than previous works and shows robustness. And finally, we discuss the effectiveness of \sys on a large scale of malicious samples.
\section*{Acknowledgment}

We sincerely thank the reviewers and anonymous shepherd for their valuable comments helping us to improve this work. This work was supported in part by grants from the Chinese National Natural
Science Foundation (61272078, 62032010, 62172201), Science Foundation for Youths of Jiangsu Province (No. BK20220772), the program B for Outstanding PhD candidate of Nanjing University.

\bibliographystyle{ieeetr}
\bibliography{ref}

\end{document}